\documentclass{aa501} 
\usepackage{graphics} 

\def\mincir{\raise -2.truept\hbox{\rlap{\hbox{$\sim$}}\raise5.truept 
\hbox{$<$}\ }} 
\def\mincireq {\hbox{\raise0.5ex\hbox{$<\lower1.06ex\hbox{$\kern-1.07em{\sim}$}$}}} 
\def\magcir{\raise -2.truept\hbox{\rlap{\hbox{$\sim$}}\raise5.truept \hbox{$>$}\ }} 
\def\gr{\kern 2pt\hbox{}^\circ{\kern -2pt K}} 
\def\e{\epsilon}

\begin{document} 
 
 
\title{X-Ray Spectral Components of Starburst Galaxies}

\author{ 
	Massimo Persic\inst{1} and 
	Yoel Rephaeli\inst{2,3}} 
 
\offprints{M.P.; e-mail: {\tt persic@ts.astro.it}} 
 
\institute{Osservatorio Astronomico di Trieste, via G.B.Tiepolo 11, 34131 
Trieste, Italy
	\and  
School of Physics and Astronomy, Tel Aviv University, Tel Aviv 69978, Israel
	\and 
CASS, University of California, San Diego, La Jolla, CA 92093, USA}
\date{Received..................; accepted...................}

\abstract{
X-ray emission processes in starburst galaxies (SBGs) are assessed, with 
the aim of identifying and characterizing the main spectral components. 
Our survey of spectral properties, complemented with a model for the 
evolution of galactic stellar populations, leads to the prediction of a 
complex spectrum. Comparing the predicted spectral properties with current 
X-ray measurements of the nearby SBGs M82 and N253, we draw the following 
tentative conclusions:
1) X-ray binaries with accreting NS are the main contributors in the 
2-15 keV band, and could be responsible for the yet uninterpreted hard 
component required to fit the observed 0.5-10 keV spectra of SBGs; 
2) diffuse thermal plasma contributes at energies $\mincir$1 keV; 
3) nonthermal emission, from Compton scattering of FIR and CMB radiation 
field photons off supernova-accelerated relativistic electrons, and AGN-like 
emission, are likely be the dominant emission at energies $\magcir$30 keV; 
4) supernova remnants make a relatively minor contribution to the X-ray 
continuum but may contribute appreciably to the Fe-K emission at 6.7 keV.
\keywords{Galaxies: X-ray -- Galaxies: spiral -- Galaxies: star formation} 
}

\maketitle 
\markboth{Persic \& Rephaeli: X-ray spectra of starburst galaxies}{}

\section{Introduction} 

In starburst galaxies (SBGs) enhanced star formation activity (lasting 
typically $\mincir 10^8$ yr) drives a chain of coupled stellar and 
interstellar (IS) phenomena that are manifested as intense far-infrared 
(FIR) and X-ray emission. The SBG class is a heterogeneous group of galaxies 
that are selected based on optical, UV and FIR properties. 
Historically, optically selected galaxies, like HII galaxies, were first 
recognized to undergo a burst of star formation (Searle et al. 1973); 
the term "starburst galaxy" was introduced by Weedman (1981; see also 
Balzano 1983). Subsequently, FIR-luminous galaxies were also recognized 
to be SBGs, a consequence of efficient heating of IS dust by the 
radiation from abundant massive stars (Soifer et al. 1986). Increased 
stellar activity leads also to a higher supernova (SN) rate, shock heating of 
IS gas, and a more efficient particle acceleration mechanism in SBGs 
compared to `normal' spirals. 

Interest in SBGs stems also from the realization that these resemble young 
galaxies in the earlier universe. Indeed, a SB phase was very common in the 
earlier universe, as the cosmic star formation rate (and hence the cosmic 
chemical enrichment) was substantially higher at epochs corresponding to 
$z \magcir 1$ (with the data being interpreted with the SFR having either a 
peak in the redshift range $1 \mincir z \mincir 2$ (Madau et al. 1996), or -- 
due to small number statistics and dust correction effects -- a plateau out 
to $z \sim 4$ (Thompson et al. 2001). So, if the main properties of SBGs in 
the present universe resemble those of normal galaxies during the 
evolutionary phase at $z > 1$, the study of local SBGs may provide insight into 
processes that occurred at that earlier epoch.

A primary manifestation of the SB activity is X-ray emission. Given the 
greatly enhanced star formation rate, energetic phenomena related to the 
final stages of stellar evolution -- X-ray binaries, supernova remnants 
(SNRs), galactic winds, and Compton scattering of ambient FIR photons off 
SN-accelerated relativistic electrons -- clearly suggest that SBGs are 
typically more powerful X-ray emitters than normal galaxies. In fact, 
normalized to the 7000 \AA{} flux (i.e., gauging activity by the old stellar 
population), in the X-ray band SBGs (as well as of other classes of active 
galaxies) are brighter than normal (spiral and elliptical) galaxies (Schmitt 
et al. 1997). The mean X-ray spectrum of SBGs is expected to reflect the 
diverse nature of high energy activity in SBGs.

The earliest attempt to determine a mean broad-band spectrum of SBGs (based 
on {\it Einstein}/IPC, {\it HEAO1}-A2, and {\it HEAO1}-A4 data for a sample 
of 51 FIR-selected putative SBGs) yielded some evidence that the (co-added) 
emission from SBGs was detectable beyond $\sim$10 keV even based on 
limited-quality survey data, and that the mean SBG spectrum was somewhat 
harder (photon spectrum index $\Gamma$ $\sim$ 1.5) than the mean AGN spectrum 
(Rephaeli et al. 1991, 1995). Individual spectral studies (based on data of 
limited spatial resolution) have shown the 0.5-10 keV spectra of SBGs to be 
complex: these are best-fit by one (or more) low-temperature ($kT < 1$ keV) 
component(s) plus a harder component, the latter interpreted as being either 
thermal with $kT \sim 5-10$ keV or nonthermal with $\Gamma \sim 1.5-2$ (M82, 
N253: Ptak et al. 1997, Cappi et al. 1999; M83: Okada et al. 1997; N2146: 
Della Ceca et al. 1999; N3256: Moran et al. 1999; N3310, N3690: Zezas 
et al. 1998). A purely thermal hard component would imply low chemical 
abundance ($Z \sim 0.3 \, Z_\odot$), whereas comparable contributions of 
thermal and nonthermal emissions would imply $Z \sim Z_\odot$ 
\footnote{Solar abundances can also be recovered using different spectral 
      models (Weaver et al. 2000). More generally, however, it should be 
      emphasized that abundance determinations are difficult due to the 
      uncertainties in the Fe-L atomic physics, because Fe-L lines couple with 
      O and Ne-K lines upon which abundance determinations rely strongly 
      (e.g., Matsushita et al. 2000).}.  
The issue of whether the hard component is actually thermal with inferred high 
temperatures and strongly subsolar abundance, and whether it originates from 
genuinely hot diffuse gas or from unresolved point sources, is not settled yet 
(e.g.: Weaver et al. 2000, Dahlem et al. 2000). Furthermore, the spectrum of 
at least some SBGs may also include a substantial contribution from a compact 
nuclear source: see, e.g., the apparent temporal variability of the 2-10 keV 
flux found in {\it ASCA} and {\it RXTE} measurements of M82 (Tsuru et al. 
1997, Ptak \& Griffiths 1999, Rephaeli \& Gruber 2001). More recently, for 
N253, {\it Chandra} data have resolved the regions where the soft thermal 
X-rays originate (Strickland et al. 2000), and {\it XMM} has separated the 
extended and point-like emission components in the disk and the nuclear 
region: the unresolved (diffuse?) emission is spatially and spectrally 
complex, with two warm plasma components in the disk and three (warm and 
hot) in the nucleus (Pietsch et al. 2001). For both N253 and M82, 
{\it Chandra} data have shown that the $\sim$ 2-10 keV flux is dominated by 
point-source emission (Strickland et al. 2000, Griffiths et al. 2000).

The main goal of this paper is an attempt to quantify the X-ray 
characteristics of the (stellar and non-stellar driven) modes of activities 
in SBGs, in order to identify basic spectral features that will help 
elucidate the nature of these galaxies. Starting from a realistic stellar 
population model for X-ray binaries and SNRs in our Galaxy, we account for 
all the viable stellar and gaseous X-ray emitting processes in a galactic 
environment, and describe the respective spectra in Section 2. In section 3 
we construct a template for the composite X-ray spectrum of a galaxy; this is 
then compared (in Section 4) with measurements of the nearby SBGs M82 and 
N253. We conclude with a summary of our main findings (Section 5).

\section{Galactic sites of X-ray emission}

X-ray emission in galaxies is either directly related to stellar activity, or 
closely associated with it through SN and SN-driven gas heating and particle 
acceleration processes. X-ray emission processes that are associated with 
stellar activity can be modelled self-consistently based on our knowledge of 
the evolution of stellar populations. Specifically, assume that any spectral 
differences among the various components stem from differences in mass, mass 
ratio, and geometry of the system. If so, the birthrate function $\nu$, the 
duration of the X-ray bright phase $\tau_{\rm x}$, the X-ray luminosity 
$L_{\rm x}$ (in a given spectral band), and the profile of the emitted 
spectrum $f(\e)$, will all be functions of the initial primary stellar 
mass $M$, the initial separation $a$, and the initial mass ratio $q$ (i.e., 
allowing for binary systems). The resulting X-ray spectrum is then 
$$
\bar f(\e) ~=~ {\int \int \int ~ \nu  \tau_{\rm x} L_{\rm x}~ f(\e) ~ 
dM \,dq\, da \over \int \int \int ~ \nu  \tau_{\rm x} L_{\rm x}~ dM \,dq\, 
da  } \,. 
\eqno(1) 
$$
As it will be discussed in the following sections, high-luminosity and 
low-luminosity X-ray binaries differ mainly in the range of $q$ values, 
whereas SN of types Ia and II differ mainly in the range of $M$ values. 
X-ray binaries span a similar mass range as SN of both types, but have very 
different periods of X-ray bright phase, as well as different spectral 
emission profiles. Based on these considerations, our aim here is to 
provide a `handbook' of the most relevant emission processes in SBGs, a 
convenient scheme for calculation of superposed spectra under various 
conditions that can be easily updated as new data become available. 
In this sense, the work reported here is a more detailed study of galactic 
X-ray emitting environments; as such, it is an extension of our previous 
work (Rephaeli et al. 1995), and more comprehensive than, e.g., the work 
of David et al. (1992) in which the respective role of various sources 
of emission in FIR-selected normal and SB galaxies was discussed on the 
energetic grounds alone.

In general, theoretical population synthesis models of SBs have been 
constructed based on the emission from individual stars (Mas-Hesse \& Kunth 
1991, Leitherer \& Heckman 1995, Meynet 1995), and also binary systems 
(Vanbeveren et al. 1997, Schaerer \& Vacca 1998, Vanbeveren et al. 1998, Van 
Bever \& Vanbeveren 1998, Van Bever et al. 1999, Mas-Hesse \& Cervi\~no 1999).

In particular, syntheses of X-ray emitting stellar populations in SBs have 
already been performed in recent years. Mas-Hesse et al. (1996), who used
evolutionary models by Mas-Hesse \& Kunth (1991) and Cervi\~no \& Mas-Hesse 
(1994), computed the multiwavelength energy distribution for two SBGs, 
including X-ray emission from massive O stars, HMXBs, and SNRs. Lipunov et 
al. (1996a) studied the temporal evolution of the HMXB binary population in 
a SB on a timescale of $10^7$ yr, and concluded that the statistics of X-ray 
binaries depend very much on the SB age. Van Bever \& Vanbeveren (2000), 
combining a close-binary population-number synthesis code with mechanisms 
of X-ray emission in young SNRs and HMXBs, studied the temporal evolution of 
the X-ray luminosity of SBs noting the importance of interacting binaries in 
the evolution and X-ray emission of SBs.

The main difference between the work reported here and previous works is 
in their respective scopes: ours is a detailed discussion of a {\it synthetic 
X-ray spectrum} of SBGs. Other differences concern: {\it (i)} the neglect, in 
earlier works, of the contributions of LMXBs, galactic winds, and Comptonized 
emission to the X-ray luminosity of SBs; {\it (ii)} the shapes of the X-ray 
spectra assumed for the various classes of sources; and {\it (iii)} the steady 
state SB considered in this work versus the evolving SBs investigated by others 
   \footnote{As for the populations of X-ray binaries observed in external 
   galaxies (see Fabbiano 1995, and Roberts \& Warwick 2000), a birth-death 
   model (Wu 2001) can calculate their luminosity functions: by modelling the 
   star formation history, i.e. by assuming a bursting (or continual) 
   formation process, the model can reproduce the presence (or absence) and 
   location of the luminosity break in the binaries' log$N$-log$S$ curves.} 
(e.g., Lipunov et al. 1996a; Van Bever \& Vanbeveren 2000).

Similarities to previous work mainly concern the overall physics underlying 
the assumed population-number synthesis models for massive stars with 
realistic frequencies of massive close binaries. Several such models exist 
(Dewey \& Cordes 1987; Meurs \& van den Heuvel 1989; Pols et al. 1991; Pols 
\& Marinus 1994; Tutukov et al. 1992; Iben et al. 1995a, 1995b; Dalton \& 
Sarazin 1995a, 1995b; Lipunov et al. 1996b; J{\o}rgensen et al. 1997) and have 
been adopted in synthetic works on stellar populations. The overall physics 
of the binary evolution scenario is similar in all models, the differences 
concerning mainly the distribution of the initial binary parameters and the 
detailed treatment of the effects of SN explosions within massive binaries.

\subsection{\bf X-ray binaries} 

Binary systems constitute the brightest class of Galactic X-ray sources 
(see reviews by White et al. 1995, and van Paradijs 1998). The primary 
factors that determine the emission properties of X-ray binaries are (1) 
nature of the accreting object, either a black hole (BH) or a neutron star 
(NS), (2) if a NS, strength and geometry of its magnetic field, (3) geometry 
of the accretion flow (disk vs. spherical accretion) from the optical 
companion, (4) mass of the accreting object, and (5) the mass accretion 
rate. Properties (1)-(3) determine the nature and location of the emission 
region (i.e., a hot accretion disk surrounding a BH, the polar cap of a NS, 
the boundary layer between the accretion disk and the NS surface), while the 
mass (of the accreting companion) and accretion rate largely determine the 
overall luminosity, spectral shape, and time variability of the emission.

An important and essentially full characterization of X-ray binaries can 
be made according to the mass of the donor star: {\it (a)} high-mass X-ray 
binaries (HMXB), where the optical component is a main-sequence star with 
$M_{\rm opt} \magcir 8 \, M_\odot$, and {\it (b)} low-mass X-ray binaries 
(LMXB), where the optical component is a post-main-sequence, 
Roche-lobe-overflowing star with $M_{\rm opt} \mincir 1 \, M_\odot$ 
   \footnote{If the optical component is a main-sequence star with 
    $1 < M/M_\odot <8$, either it does not support strong radiatively driven 
    winds, and/or the common binary envelope (which forms when the accretion 
    rate onto the compact component is so large that $L_{\rm accr} > 
    L_{\rm Edd}$; see below) is very short lived, so that the  
    ensuing X-ray emission is unlikely to be observed.}.
The main distinction between HMXBs and LMXBs has to do with the nature of the 
mass transfer -- inefficient wind accretion in HMXBs, and efficient transfer 
through Roche-lobe throat in LMXBs [as a consequence, generally, 
$L_{\rm x}^{\rm H} \mincir {\cal O} (-1) \, L_{\rm x}^{\rm L}$]. 

An alternative and equally effective distinction, partially overlapping 
with the previous one, can be made by the nature of the accreting object: 

\noindent
{\it (a)} systems containing a strongly magnetized neutron star (NS): these 
systems are HMXBs;

\noindent
{\it (b)} systems containing a weakly magnetized NS: these systems are LMXBs; 
and 

\noindent
{\it (c)} systems containing a black hole (BH): these systems can be both 
HMXBs and LMXBs. 
\noindent

The following brief comments should be noted (see van Paradijs 1998).

Virtually all HMXBs harbor strongly magnetized X-ray pulsars 
($B \magcir 10^{12}$ G, as also inferred from the presence of cyclotron 
lines in their X-ray spectra; see Mihara et al. 1991 and references 
therein); on the other hand, X-ray pulsations occur only rarely in LMXBs, 
while bursts 
    \footnote{Luminosity increases by a factor $\magcir$10 usually within 
    1 s, followed by a decay to pre-burst X-ray flux level within 10 s to a 
    minute.} 
do occur only in LMXBs. The mutual exclusion of pulsations and bursts (but see 
Kouveliotou et al. 1996) suggests that the feature distinguishing the NSs in 
LMXBs from those in HMXBs is either a weaker magnetic field, or the alignment 
of the NS magnetic and rotational axes. However, the difference in the X-ray 
spectral properties of HMXBs and LMXBs (the former usually having harder 
spectra, see White \& Marshall 1984) strongly points to a difference in the 
geometry of the accretion flow, and hence of the magnetic field strength.
    \footnote{The difference in magnetic field strength between NSs in 
    HMXBs and those in LMXBs could be either primordial or evolutionary. In 
    the former case it may be related to a difference in the formation 
    mechanisms of NSs in HMXBs and in LMXBs, i.e.,
    via the normal evolution of a massive star and via the accretion-induced 
    collapse of a white dwarf, respectively. In the latter case, a decay of 
    the NS magnetic field -- either spontaneous or caused by (e.g.) the 
    accretion process -- has been suggested by the observation that (strongly 
    magnetized) NSs in HMXBs are all young objects, whereas those (weakly 
    magnetized) in LMXBs are typically much older.} 
In fact, in a NS with magnetic fields of $\sim 10^{12}$ G (and sub-Eddington 
accretion rates), the accretion flow is disrupted at several hundred NS radii 
and most of the inflowing material is funneled onto the magnetic pole and 
reaches the NS on a relatively small area near the magnetic polar cap. The 
emission is magnetically beamed (either along or perpendicular to the field 
lines, corresponding to, respectively, "pencil beam" or "fan beam" emission) 
and hence, if the magnetic and rotation axes are misaligned and if the 
beamed emission from the magnetic poles rotates through the line of sight to 
the observer, X-ray pulsations are observed; the emitted X-ray spectrum has a 
broken-power-law profile. (Note that these features collectively correspond 
to overall HMXB phenomenology.) On the other hand, for much weaker ($ < 10^9$ 
G) magnetic fields, the accretion disk may touch or come close to the NS 
surface, and the accreting material is distributed over a larger fraction of 
the NS surface: consequently, X-ray emission shows no pulsations (but it may 
show bursts 
     \footnote{After sufficiently large amount of 
     matter has accreted on the NS surface, critical conditions may develop at 
     the base of the accreted flow, causing unstable helium burning: the 
     sudden release of nuclear energy gives rise to an X-ray burst.}
), and has a partially Comptonized thermal spectrum $\propto \e^{-\Gamma} 
e^{-\e/kT}$, with $\Gamma \sim 1.0-1.4$. (These features collectively 
correspond to overall LMXB phenomenology.) 

Spectra of BH X-ray binaries (BHXBs) have a distinct two-component signature 
(which has turned out to be a very good predictor for the presence of a BH 
in an X-ray binary; e.g., Tanaka \& Lewin 1995). One component is {\it 
ultrasoft} with a characteristic temperature of $\magcir$1 keV, which 
can be modelled as an optically thick, geometrically thin accretion disk 
(e.g., Ebisawa et al. 1994). The other component is an {\it ultrahard} power 
law with photon index in the range $\sim$1.5 to $\sim$2.5 that extends up to 
several hundred keV (e.g., Wilson \& Rothschild 1983). There is a correlation 
between luminosity and shape of the 2-10 keV spectrum: a spectral hardening, 
caused by the weakening of the ultrasoft component, signals a transition from 
a high state, where the ultrasoft component may even completely dominate the 
emission, to a low state, where the ultrasoft component is weak or absent, 
and the X-ray spectrum is dominated by the hard power-law component. Out of 
10 BHXBs currently known in the Galaxy 
    \footnote{Other 17 systems are suspected to be BHXBs based on spectral 
    considerations (see van Paradijs 1998).}, 
3 are HMXBs, and 7 are transient LMXBs (White \& van Paradijs 1996). The total 
number of BH X-ray transients is estimated to be $N_{_{\rm BHXT}} \sim$500, 
with typical recurrence time $\Delta t_{\rm r} \sim 100$ yr, and decay time 
$\tau_{\rm d} \sim 0.5$ yr (White \& van Paradijs 1996). This implies that at 
any given time only $N_{_{\rm BHXT}} \tau_{\rm d} / \Delta t_{\rm r} \sim 2-3$ 
X-ray active BHXBs are expected in the Galaxy. Given the relative paucity of 
such sources within the two main families of X-ray binaries, the contribution 
from BHXBs will not be considered further in this paper. 

We adopt the model of Iben et al. (1995a,b) for the formation of the Galactic 
X-ray binary population. The model is based on standard stellar evolution 
theory, and a numerical code with a semi-empirical birth function for binary 
stars (normalized to parameters appropriate for the Galaxy),
$$
d^3\nu \,{\rm (yr^{-1})}~=~0.2~d{\rm log}a~\biggl({dM \over M^{2.5}}\biggr) 
~dq\,,
\eqno(2)
$$
where $a$ is the semimajor orbital axis of the binary orbit (in the range 
$10 \mincir a/R_\odot \mincir 10^4$), $M$ is the mass of the primary, and 
($q \leq 1$) is the secondary-to-primary mass ratio (all quantities refer to 
the primordial system). The Iben et al. model is based on the assumption 
of equal binary birth and death rates and the attainment of steady state, 
valid when the age of the stellar population is much longer than the 
characteristic system lifetime.

\subsubsection{\it HMXBs}

Persistent HMXBs have short periods ($P < 10$ days) and show significant 
flux variability; these objects form in systems with $11.4 \leq M / M_\odot 
\leq 50$ and $10 \leq qM / M_\odot \leq 30$, and $a < 2000 \, R_\odot$ (so 
the primary star can make Roche-lobe contact when it leaves the main 
sequence). Transient HMXBs have recurrent pronounced flux spikes 
and periods of quiescence where their X-ray emission is below the limit of 
detectability; these objects have long periods ($P > 10$ days) and form in 
systems with $q \simeq 1$. Conservative (binary) mass transfer 
increases the secondary mass and the separation. A NS is formed as the 
remnant of a SN explosion of the primary star with mass $11.4 <M /M_\odot 
<40$. (The lower limit can be $10 \, M_\odot$ for binaries wide enough that 
the primary never fills the Roche lobe, but such systems do not evolve into 
X-ray binaries.) A $>$40$ \, M_\odot$ (primary) star is assumed to evolve 
into a $\sim$10$\, M_\odot$ BH.

X-rays from HMXBs result from accretion by a NS of matter from the 
radiatively driven wind of an OB star. The bright X-ray phase of persistent 
HMXBs begins when the optical component approaches the Roche lobe in size and 
the fraction of the donor-emitted wind matter captured by the NS (or BH) 
increases towards maximum. In order to estimate the birthrate of binary 
systems in the correct configuration, Iben et al. (1995a) assume 
that the accretion occurs at the Bondi \& Hoyle (1944) rate, and that all the 
gravitational potential energy released in the accretion process in converted 
to X-rays. The luminosity is 
$$
{L_{\rm x} \over L_\odot} ~\sim~ {4 \times 10^{11} \over \alpha_w^4} 
\biggl({M_{\rm x}  \over M_{\rm opt}} \biggr) \biggl({R_{\rm opt} 
\over a}\biggr)^2 \dot M_w \,,
\eqno(3a)
$$
where $M_{\rm opt}$ and $R_{\rm opt}$ are the donor (OB star) mass and radius, 
$M_{\rm x}$ is the primary mass, $\dot M_w$ is the mass-loss rate from the 
donor, and $\alpha_w \sim 1-R_{\rm opt}/a$ (Waters \& van Kerkwijk 1989). 
Choosing the fairly typical values $M_{\rm x} \sim 1.4\,M_\odot$, 
$M_{\rm opt} \sim 14\, M_\odot$, $R_{\rm opt}/a \sim 0.6$, and $\dot M_w 
\sim 10^{-6} M_\odot$ yr$^{-1}$ (see Iben et al. 1995a), we obtain: 
$$
L_{\rm x} ~\sim~ {1500 \over \alpha_w^4} L_\odot ~ \sim~ {6 \times 10^{36} 
\over \alpha_w^4} ~~ {\rm erg~ s}^{-1}\,.
\eqno(3b)
$$
Thus, to achieve the highest possible $L_{\rm x}$ and persistence of the 
source, the optical component must be close to filling its Roche lobe 
($R_{\rm opt} \rightarrow a$). In this case the gas stream dominates the 
accretion flow and a stable accretion disk forms around the primary, 
smoothing out inhomogeneities in the accretion and reducing the time 
variability of the source. 

Using $\Delta{\rm log}a \simeq 2$, $M =(11.4-50) \, M_\odot$, and $\Delta q 
\sim 0.3$ in eq.(2), so as to obtain a crude estimate of the birthrate of 
systems having a NS and an OB star in the correct configuration, Iben et al. 
(1995a) estimate $\nu_{_{\rm H}} ~\sim~ 2 \times 10^{-3}~~ {\rm yr}^{-1}$. 
Since the average duration of the bright X-ray phase ($L_{\rm x} >10^{37}$ 
erg s$^{-1}$) is $\tau^{^{\rm H}}_{\rm x} \sim 2.5 \times 10^4$ yr (Meurs \& 
van den Heuvel 1989), the expected number of such systems that are produced 
during the massive star formation phase, $N_{_{\rm H}} = \nu_{_{\rm H}} 
\tau^{^{\rm H}}_{\rm x}$, is 
$$
N_{_{\rm H}} ~\sim ~50\,,
\eqno(4)
$$ 
in agreement with observational estimates. 

The phase-averaged X-ray photon spectrum from persistent, bright 
($L_{\rm x} \magcir 10^{37}$ erg s$^{-1}$) HMXBs in the 2-50 keV band 
can be represented (White et al. 1983) as a broken power law
     \footnote{This spectral steepening has been interpreted as the result 
     of the anisotropic energy dependence of the cross sections below the 
     cyclotron resonance in strong ($>$10$^{12}$ G) magnetic fields (e.g., 
     Boldt et al. 1976), and it may give an independent measure of the field
     strength. But other parameters, e.g. the temperature and optical depth of 
     the NS atmosphere, could be important in determining the break energy.}:
$$
f^{^{\rm H}}(\e) ~ \propto ~ 
\left\{
\begin{array}{ll} \e^{-\gamma}  
& \mbox{ $~.~.~.~$ if $\e \leq \e_c$} \\
\e^{-\gamma}~ e^{-[(\e-\e_c) / \e_F]}   
& \mbox{ $~.~.~.~$ if $\e > \e_c$ }\,,
\end{array} 
\right.
\eqno(5)
$$
where $\e$ denotes energy (in keV), and with photon index $\gamma \simeq 
1.2^{+0.2}_{-0.2}$, cutoff energy $\e_c \sim 20$ keV, and $e$-folding 
energy $\e_F \sim 12$ keV. In addition, a small amount of low-energy 
absorption by cool material and an Fe-K feature between 6 and 7 keV are 
required in the fits. There appears to be no correlation of $\gamma$, $\e_c$, 
or Fe-K equivalent width with luminosity. 

The mean HMXB spectrum is then formally obtained by integrating $f^{^{\rm H}}$ 
over the HMXB birthrate function: 
$$
\bar J_{_{\rm H}}(\e) ~= ~{\int \int \int_{11.4}^{50.0} \nu \tau_{\rm x} 
L_{\rm x}\, f^{^{\rm H}}(\e) \, dM\, da \, dq
\over    
\int \int \int_{11.4}^{50.0}  \nu \tau_{\rm x} L_{\rm x} ~ dM \, da \, dq } \,.
\eqno(6)
$$
Because there appears to be no obvious correlation between X-ray spectral 
shape and mass ratio $q$, primary mass $M$, and/or separation $a$ (while 
the system must be close, $a < 2\,R_{\rm opt}$, in order for $L_{\rm x} 
\geq 10^{37}$ erg s$^{-1}$), eq.(6) reduces to:
$$
\bar J_{_{\rm H}}(\e) ~=~ f^{^{\rm H}}(\e)\,.
\eqno(7)
$$

A striking characteristic in the HMXB spectrum is a prominent Fe-K emission 
feature with central energy in the range $6.4 - 6.7$ keV, and 
equivalent width $EW_{\rm Fe-K} \sim 0.2-0.6$ keV, only weakly dependent on 
$L_{\rm x}$ (White et al. 1983). This emission is unlikely to come from the 
region close to the NS surface, because the line would be gravitationally 
redshifted to $\sim 5.5-5.8$ keV. It is also unlikely that the line is 
produced near the photosphere of the optical companion, or from the accretion 
disk surrounding the NS, because given the small angular sizes involved, the 
expected contributions to fluorescent emission should be both too modest and 
dependent on the orbital phase, contrary to observations. The two remaining 
potential sites for fluorescent emission to occur are the stellar wind of the 
optical companion and the magnetosphere of the NS (see Pravdo et al. 1977; 
White et al. 1983; Ebisawa et al. 1996). Another relevant emission feature 
observed in HMXB spectra (at least when the NS companion is eclipsed) is due 
to S{\tt XV} (Sako et al. 1999; Ebisawa et al. 1996; Nagase et al. 1994).

\subsubsection{\it LMXBs}

Galactic LMXBs are divided into bulge and disk populations. It is estimated 
that a few hundred LMXBs exist in the Galactic disk, of which only $\sim$100 
are persistent X-ray sources with $L_{\rm x} \sim L_{\rm Edd} \sim10^{38}$ 
erg s$^{-1}$, where $L_{\rm Edd}$ is the Eddington limit for a NS accretor. 

In a LMXB system the secondary is a low mass star whose enevelope fills the 
Roche lobe. Mass transfer onto the NS or BH primary is driven by evolutionary 
expansion of the donor, magnetic stellar wind, or gravitational radiation. 
Based on the semi-empirical birth function of such systems, Iben et al. 
(1995b) estimated the Galactic disk LMXB birthrate to be $\nu_{_{\rm L}} 
\sim (1-4) \times 10^{-5}$ yr$^{-1}$. 

Iben et al. (1995b) treat the modes of conservative mass transfer from a 
Roche-lobe--filling (optical) secondary onto the (degenerate) primary in the 
approximation that the energy from the accretor, intercepted by the donor, 
has no feedback on the mass-loss rate from the donor itself. This discussion 
applies both to the case of LMXBs resulting from the evolution of massive 
binaries with a large $q$ (in which the accretor is a NS or a BH, which 
results from an initial primary mass in the range $11.4 \leq M/M_\odot \leq 
50$), and to the case of LMXBs resulting from accretion-induced collapse of 
(primary) O-Ne white dwarfs (which results from initial primary masses in the 
range $9 \leq M/M_\odot \leq 11.4$). Iben et al. (1995b) then argue that the 
predicted and the observed Galactic populations of LMXBs can be reconciled 
only if the assumptions of conservative mass transfer and of 
no-accretor/donor-feedback are relaxed. In fact, an irradiation-induced wind 
will develop from the donor that can remove from the system an 
order of magnitude more mass than it is actually tranferred to the accretor. 
Also, the induced stellar wind can be accreted by the NS, a bootstrapping 
situation can be established, in which accretion from an induced wind emitted 
by the optical component that does not fill the Roche lobe is sufficient for 
producing the radiation necessary to produce the induced wind itself (the 
feedback loop is stable). These arguments lead to an estimated duration of 
the X-ray bright stage of $\tau^{^{\rm L}}_{\rm x} \sim (0.5-1) \times 10^7$ 
yr which in turn, being $N_{_{\rm L}} = \nu_{_{\rm L}} 
\tau^{^{\rm L}}_{\rm x}$, leads to 
$$
N_{_{\rm L}} ~\sim ~ 200\,,
\eqno(8)
$$ 
in fair agreement with observations.

The interpretation of LMXBs spectra has progressed with increasing 
quality of available data, in terms of: optically thin thermal bremsstrahlung 
(TB), blackbody (BB) plus TB, BB plus an optically thick (but physically thin) 
accretion disk blackbody (DBB), and BB plus a form that could possibly 
represent a Comptonized spectrum (see Christian \& Swank 1997 for references). 
The ambiguity in the spectral form of LMXBs is partly due to the relatively 
narrow detector spectral band (as compared to the respective emission 
spectrum). Following White et al. (1985), Christian \& Swank (1997) have 
shown that a model of the form $\e^{-\Gamma}$exp($-\e/kT$) [the so-called 
"unsaturated Comptonization" (USC) model 
    	\footnote{In accretion onto a NS or a BH the simple BB 
	accretion disk model is not correct and it must be modified 
	to account for the effects of electron scattering, which make 
	energy losses by Comptonization of cool photons on hot electrons 
	dominate the spectral formation (Sunyaev \& Titarchuk 1980; see 
	also White et al. 1988). Starting from thermal bremsstrahlung, 
	$f(\e) \propto g(\e) \, e^{-\e/kT}$ with a Gaunt factor $g(\e) 
	\sim (\e/kT)^{-1.3}$ in the region of interest, White et al. 
	(1985) write, in more generality, $f(\e) \propto \e^{-\Gamma} 
	e^{-\e/kT}$ which approximates the unsaturated Comptonized 
	spectrum of cool photons upscattered on hot electrons (see also 
	Treves et al. 1988).}] 
is a good approximation to the 0.5 - 20.0 keV {\it Einstein}/SSS+MPC spectra 
of Galactic-disk LMXBs. The USC model gives the best fits to the data, and 
although its results may not be directly interpretable in terms of a physical 
description of the sources,
      \footnote{E.g., the USC model can approximate TB if $\Gamma 
	\sim 1.2 - 1.4$ [the Gaunt factor for energy spectrum being $g(\e) 
	\simeq \e^{-0.29}$ in the 2-10 keV range], or BB+TB if $\Gamma 
	\sim 1.0 - 1.4$.}, 
it does offer a phenomenological description with which to compare them. 
Sources showing variations in intensity by a factor of $\sim$2 ("dippers") 
or larger ("bursters", some of which show periodic or aperiodic dips or 
eclipsing behavior -- see Christian \& Swank 1997 and references therein), 
can be fitted with the USC model both in their high (bright) states and in 
their low (faint) states, with the spectral parameters changing accordingly: 
higher luminosities tend to correspond to lower values of $\Gamma$ and 
viceversa (Christian \& Swank 1997; the trend also applies to nonvariable 
sources). Qualitatively, this inverse correlation can be understood within 
the accretion model devised to explain the high X-ray luminosities of LMXBs: 
the increase in luminosity is driven by an increase of the mass accretion 
rate $\dot M$, which in turn means a piling up of material around the 
emission region (e.g., the magnetic polar cap of the NS) that leads to a 
higher Compton scattering optical depth, and hence to a lower value of 
$\Gamma$ 
    	\footnote{In the USC model the relation linking the 
	spectral index $\Gamma$ to the scattering optical depth 
	$\tau$ is 
	$$
	\Gamma ~=~ \sqrt{ {9 \over 4} ~+~ \biggl( 
	{1.68 \times 10^3 \over (kT/{\rm keV}) } \biggr) ~ \biggl
	(\tau + {2 \over 3}\biggr)^{-2} } ~-~ {1 \over 2} 
	$$ 
	(Sunyaev \& Titarchuk 1980). Note that for sources having 
	$\Gamma <0.5$, i.e. sources that are almost optically 
	thick, the unsaturated Comptonization approximation is not 
	strictly valid, but the USC model still provides the best 
	fit to the data (Christian \& Swank 1997).}. 
The overal spectral properties of LMXBs can be broadly divided into 
two classes based on whether the overall luminosity of the source is 
lower or larger than $\sim 10^{37}$ erg s$^{-1}$ (White et al. 1988). In 
particular, the X-ray spectra of Galactic-disk LMXBs in the 0.5-20 keV 
band can be described by: 
$$
f^{^{\rm L}}(\e) ~ \propto ~
\left\{
\begin{array}{ll} ~ \e^{-1.4}~ e^{-\e/kT} 
& \mbox{$~.~.~.~$ if $L_{\rm x} < 10^{37} {\rm erg \,s}^{-1} ~ ~ (9a)$} \\
~ e^{-\e/kT} 
& \mbox{$~.~.~.~$ if $L_{\rm x} \ge 10^{37} {\rm erg \,s}^{-1}  ~ ~ (9b) $} 
\end{array} 
\right.
$$
with $5 \mincir kT \mincir 10$, and both $\e$ and $kT$ in keV (see 
Christian \& Swank 1997). [As the ratio of the total energy emitted 
in bursts to that emitted in persistent flux is ${\cal O}(-2)$, for our 
purposes here we assume that a given LMXB system can be uniquely 
described by eq.(9a) or eq.(9b).] The LMXB populations of globular 
clusters in the Galactic bulge and in M31 have similar characteristics 
to those in the Galactic disk (Callanan et al. 1995; Matsushita et al. 
1994; Trinchieri et al. 1999). Thus, the spectral properties of LMXBs 
vary little (if at all) between the disk and the globular-cluster 
populations of our Galaxy, or among galaxies in general.

The mean LMXB spectrum is then obtained by integrating $f^{^{\rm L}}(\e)$ 
over the LMXB birthrate function:
$$
\bar J_{_{\rm L}}(\e)  ~=~ {
\int \int \int_{9.4}^{50.0} \nu \tau_{\rm x} L_{\rm x} ~
f^{^{\rm L}}(\e) ~ dM\, da \, dq
\over    
\int \int \int_{9.4}^{50.0}  \nu \tau_{\rm x} L_{\rm x} ~ dM \, da \, dq } \,.
\eqno(10)
$$
Assuming that $f^{^{\rm L}}$ is independent of both $M$ and $q$, the 
resulting average LMXB spectrum is the sum of the low-$L_{\rm x}$ and 
high-$L_{\rm x}$ contributions [eq.(9)], weighted by the relative 
frequencies. We can assume that the most critical parameter in determining 
whether $L_{\rm x}$ is higher or lower than the "threshold" value of 
$10^{37}$ erg s$^{-1}$ is the initial semimajor axis $a$. In the above 
model (sect. 2.1.1), tight (2 $\mincir a/R_{\rm opt} \mincir$ 8) systems 
have $L_{\rm x} \magcir 10^{37}$ erg s$^{-1}$, while loose 
(8 $\mincir a/R_{\rm opt} \mincir$ 100) systems have $L_{\rm x} \mincir 
10^{37}$ erg s$^{-1}$. Thus, eq.(10) transforms into
$$
\bar J_{_{\rm L}}(\e)  ~=~ 
0.65\, f^{^{\rm L}}_{{\rm lo-}L_{\rm x}}(\e) \, + \, 
0.35\, f^{^{\rm L}}_{{\rm hi-}L_{\rm x}}(\e) 
\eqno(11)
$$
where $f^{^{\rm L}}_{{\rm lo-}L_{\rm x}}$ and 
$f^{^{\rm L}}_{{\rm hi-}L_{\rm x}}$ denote the low-$L_{\rm x}$ and 
high-$L_{\rm x}$ representations of eq.(9).

\subsection{\bf Supernova remnants}

It is usually assumed that all stars with mass between some 
lower threshold and $40 M_\odot$ will eventually explode as supernovae 
(e.g., Woosley \& Weaver 1995). (The evolution of $>40 M_\odot$ stars is 
less clear.) The lower mass for a progenitor of a SN II event (core collapse) 
is $8 \, M_\odot$ (if convective overshooting is not important; if it is, the 
lower limit could be as low as $5\, M_\odot$). According to the standard 
model, SN Ia is a result of thermonuclear instability in a degenerate C-O 
white dwarf that ignites owing to thermal runaway when the mass reaches 
the Chandrasekhar limit, $M_{\rm Ch} \simeq 1.4\, M_\odot$ (the exact value 
depends on chemical composition), triggered by the accumulation of material 
from a companion star.

X-ray emission from SNRs occurs mostly during phase 1 (free expansion) and 
part of phase 2 (adiabatic) of the evolution of SNRs (e.g.: Woltjer 1972, 
Chevalier 1977), for a typical duration of $\tau^{^{\rm SNR}}_{\rm x} 
\mincir 10^3$ yr. The spectrum can be generally described as thermal 
     \footnote {The ionization equilibrium in the plasma is an additional 
     variable. In detail, individual SNRs may exhibit non-thermal emission 
     in addition (e.g. SN1006: Vink et al. 2000) to or instead of the thermal 
     component (e.g. G266.2-1.2: Slane et al. 2001). Spatially resolved data 
     show that SNR spectra are spatially complex (e.g., RCW 86: Borkowski et 
     al. 2001b, Bocchino et al. 2000; for a review see Hwang 2001).}
with: $kT \sim 2  ~ {\rm keV}$, $Z \sim Z_\odot$. The X-ray luminosities of 
young SNRs are typically $L_{\rm x} \sim 10^{37}~ {\rm erg ~s}^{-1}$ (e.g.: 
Hamilton \& Sarazin 1984, Charles \& Seward 1995, Burrows et al. 2000, 
Borkowski et al. 2001a). Since the estimated Galactic SN birthrate is 
$\nu_{_{\rm SN}} \simeq 2 \times 10^{-2}$ yr$^{-1}$, the predicted number of 
X-ray bright SNRs (i.e., young remnants that are in the ejecta dominated 
phase, like Cas A or Tycho's SNR) in the Galaxy, 
$N_{_{\rm SNR}} = \nu_{_{\rm SN}} \tau^{^{\rm SNR}}_{\rm x}$, is 
$$
N_{_{\rm SNR}} \sim 20\,,
\eqno(12)
$$
in fair agreement with observations
      \footnote{A catalog, based on {\it Chandra} data and complete 
      down to $L_{\rm 0.125-8.0~keV} \sim 10^{36}$ erg s$^{-1}$, of 110 
      sources located in a large central region of the nearby face-on 
      spiral M101 -- a system that is considered to be in many ways similar 
      to the Galaxy -- contains 9 SNRs (Pence et al. 2001). The implied 
      relative frequency of SNRs among the detected sources, 0.080, is in 
	substantial agreement with the value, $\sim$0.074, deduced for our 
	model.}.
[Later in their evolution SNRs cool and dim: during most of the second 
(adiabatic) phase, remnants are likely to have lower temperatures, $0.5 
\mincir kT/{\rm keV} \mincir 1$ keV; in the third (radiative) phase, 
remnants will cool rapidly to temperatures too low for X-ray emission.]

The resulting average SNR spectrum is then:
$$
\bar J_{_{\rm SNR}} ~=~ {
\int_{9.4}^{50} \nu \tau_{\rm x} L_{\rm x} J_{_{\rm II}} dM  \, + \, 
\int_{5}^{9.4} \nu \tau_{\rm x} L_{\rm x} J_{_{\rm Ia}} dM 
\over    
\int_{9.4}^{50} \nu \tau_{\rm x} L_{\rm x}\, dM  \, + \, 
\int_{5}^{9.4} \nu \tau_{\rm x}  L_{\rm x}\, dM 
}\,,
\eqno(13)
$$
where $J_{_{\rm Ia}}$ and $J_{_{\rm II}}$ denote, respectively, the average 
spectra of the remnants of type Ia and type II supernovae (which in principle 
may differ from each other). The RHS of eq.(14) accounts for: {\it (a)} SNII 
explosions of (all) stars with $9.4 \leq M/M_\odot \leq 11.4$ (leading to a 
C-O white dwarf remnant), of (all) stars with $11.4 \leq M/ M_\odot \leq 40$ 
(leading to a NS remnant), and of (all) stars with $40 \leq M/M_\odot \leq 
50$ (leading to a BH remnant); and {\it (b)} SN Ia explosions from accreting 
white dwarfs in binary systems, with the white dwarfs resulting from the 
evolution of progenitor stars with $5 \leq M/ M_\odot \leq 9.4$ 
      \footnote{The lower mass limit is determined by the condition that a 
      main-sequence star becomes a white dwarf within the lifetime of the SB 
      ($\sim$10$^8$ yr). Generally, evolution times are longer for lower 
      masses (e.g., Maeder \& Meynet 1989).}.
If the expanding SNRs have all similar spectra, irrespective of the type of 
the SN event and of mass, then 
$$
\bar J_{_{\rm SNR}} ~\propto~ j_{_{\rm th}}(\e; \, kT=2 \,{\rm keV}, \, 
Z=Z_\odot)
\eqno(14)
$$ 
where $j_{_{\rm th}}$ is the thermal bremsstrahlung emissivity function.

\subsection{\bf O stars}

It is well known that massive O and early-B stars are sources of X-ray 
emission (Rosner et al. 1985). If about $50\%$ of O stars are in 
binary systems (e.g., Garmany et al. 1980), then based on the birth rate of 
HMBXs (see section 2.1.1) we have $\nu_{_{\rm O \, stars}} ~\sim~ 4 \times 
10^{-3}$ yr$^{-1}$. Since the main-sequence lifetime $\tau_{\rm MS}$ of a 
$(10-30)\,M_\odot$ star (mass range appropriate for O stars) is $\tau_{_{\rm 
MS}} \sim (3-10) \times 10^6$ yr (Chiosi \& Maeder 1986; Maeder \& Meynet 
1989), the expected number of O stars is $N_{_{\rm O \, stars}} ~\sim ~2 
\times 10^4$. (This may be an upper limit: catalogs of Galactic O stars 
contain $\sim$1000 objects, see Conti \& Underhill 1988). If we assume an 
individual X-ray luminosity of $L_{\rm x} \sim 3 \times 10^{32-33}$ 
erg s$^{-1}$ (see Corcoran et al. 1994, Griffiths et al. 2000), the 
inferred total X-ray luminosity of the O-star population is $L_{\rm x} 
\sim 6 \times 10^{36-37}$ erg s$^{-1}$. The spectra of O stars have been 
successfully fitted with multi-temperature thermal models: e.g., simultaneous 
fits to the {\it ASCA}/SIS0 and {\it ROSAT}/PSPC spectra of $\delta$ Ori 
require (at least) three components with temperatures near 0.1, 0.3, and 0.6 
keV, plus absorption by a circumstellar medium, presumably the stellar wind 
(see Corcoran et al. 1994). 

The inferred X-ray faintness of O-stars (their totel luminosity would be 
matched, or even exceeded, by that of a single typical HMXB), leads to the 
realization that O stars may not contribute significantly to the X-ray 
emission of a (starbursting) stellar population
      \footnote{In M82, {\it Chandra} observations have revealed a 
      number of bright OB associations, the brightest of which has an X-ray 
      luminosity of $\sim 5 \times 10^{38}$: this large X-ray emission, 
      however, presumably arises from a combination of diffuse plasma, 
      unresolved HMXBs, and O stars (Griffiths et al. 2000). }.
Hence the contribution from O stars will not be considered further in this 
paper.

\subsection{\bf Compton scattering}

As has already been mentioned, the high SN rate in a SBG is bound to yield 
high relativistic electron densities since SN shocks are known to be primary 
sites of cosmic ray acceleration. This means that we should expect the mean 
relativistic electron density in a SBG to be much higher than in the Galaxy. 
Moreover, the mean energy density of a SBG with a FIR luminosity of $10^{11} 
L_{\odot}$ is $\sim 4\times 10^{-12}$ ergs cm$^{-3}$ (within 10 kpc radius), 
i.e. $\sim 10$ times more intense than the cosmic microwave background (CMB) 
radiation. Obviously, the more spatially concentrated the FIR emission, the 
higher is this energy density. Radiative losses dominate at electron energies 
above few hundred MeV (Rephaeli 1979), and Compton scattering of a $100 \mu$ 
photon by a $\sim$1 GeV electron boosts the energy of the photon to $\sim$10 
keV. To fully account for a luminosity of $10^{40}$ erg s$^{-1}$, the energy 
density in electrons has to be $\sim 2\times 10^{-13}$ ergs cm$^{-3}$, i.e. 
$\sim 10$ times higher than in the Galaxy. The latter value also corresponds 
(roughly) the SBG/Galaxy X-ray luminosity ratio. (Clearly, the required 
relativistic electron energy density is lower if only part of the total 
emission is due to Compton scattering, or if the FIR emission is centrally 
concentrated.)

A high density of electrons in SBGs would also yield an enhanced radio 
emission. Indeed, there exists a definite correlation between the radio 
($L_R$) and FIR luminosities of spiral galaxies: in the FIR luminosity range 
relevant to SBGs, $L_R \propto L_{FIR}^{1.3}$ (Wunderlich \& Klein 1988). 
There are several studies of SBGs as a class in the radio (e.g.: Garwood et 
al. 1987, Condon et al. 1990 and 1991, Wang \& Helou 1992): of the nearby 
SBGs, M82 in particular has been studied in detail (e.g.: Kronberg et al. 
1985, Klein et al. 1988) and is known to have an extended region of disk 
and halo emission.

If the same electron population produces both the radio and hard X-ray 
emission, then both spectra are power-laws with {\it roughly} the same 
index. The exact relations between the radio and Compton X-ray flux have 
been written down (Rephaeli 1979) and explored in detail in the context of 
SBGs by Goldshmidt \& Rephaeli (1995); a power-law spectrum of the form 
$$
f_{_{\rm C}}(\e) ~=~ K_{_{\rm C}} \, \e^{-\beta}
\eqno(15)
$$
is expected, with (a photon) index $\beta$ which is roughly in the range 
$\sim 1.6 - 1.8$. The coefficient $K_{_{\rm C}}$ can be expressed in terms 
the measured radio flux and mean value of the magnetic field in the 
emitting region.

\subsection{\bf Diffuse thermal plasma}

The ISM and the galactic-halo gas are expected to be shock-heated by SN 
explosions to approximately galactic virial temperatures, $kT \mincir 1$ keV. 
Thermal soft X-ray  emission is then expected from the gas. Indeed, 
{\it ASCA} and {\it BeppoSAX} data have systematically and unambiguously 
revealed the presence of a $<$1 keV thermal component in the 0.5-10 keV 
spectra of SBGs (Ptak et al. 1997, Cappi et al. 1999, Okada et al. 1997, 
Della Ceca et al. 1999, Moran et al. 1999, Zezas et al. 1998). More 
specifically, for N253 {\it Chandra} data have shown that soft thermal X-rays 
come from the regions of interaction between the fast SB-driven wind and the 
denser ambient ISM, not from the wind fluid itself (Strickland et al. 2000). 
The spatial resolution attained with {\it XMM} has allowed separating the 
extended and point-like emission components in the disk and the nuclear 
region, showing an ever increasing spatial and spectral complexity of the 
unresolved (diffuse?) emission. Two thin plasma components 
(with $kT \sim 0.13$ and 0.4 keV) are required in the disk, and three 
(with $kT \sim$ 0.6, 0.9, and 6 keV) in the nucleus (Pietsch et al. 2001). 
These results are consistent with the results of Strickland \& Stevens's 
(2000) simulations of SB-driven galactic winds, where the soft X-rays come 
from the region of wind/ISM interaction, which is characterized by a 
multi-temperature, non-uniform plasma. Galactic winds are more efficient at 
carrying the SN-synthesized metals, rather than the unprocessed gas, out of 
the galaxy. Since SN Ia products have the largest ejection efficiency (more 
so than SN II products) and SN Ia produce a substantial fraction of Fe, 
$\alpha$-burning to Fe ratios are predicted to be higher for the ambient 
disk gas than for the wind-borne material (Recchi et al. 2001).

The $\magcir$2 keV thermal emission from the galactic wind comes mostly from 
the central SB region itself (see Pietsch et al. 2001 for {\it XMM} data on 
N253). In Strickland \& Stevens's (2000) simulations the relatively small 
volume and the high gas density of the SB region explain why the hard X-ray 
luminosities are typically ${\cal O} (-2)$ of the soft X-ray luminosity of 
the wind (see also Suchkov et al. 1994). Note that {\it Chandra} data have 
shown that in N253 the hard ($2-8$ keV) flux is dominated by previously 
unresolved point-source emission (Strickland et al. 2000), while in M82 up 
to $75\%$ of the 2-10 keV flux is resolved into point sources (Griffiths et 
al. 2000). 

An appropriate description of the X-ray emission from thermal plasma in SBGs 
is then: 
$$
f_{\rm g}(\e) ~=~ K_{\rm g} \times 
                      j_{_{\rm th}}(\e; ~kT=0.7 \,{\rm keV}, \,Z=Z_\odot) 
\eqno(16)
$$
where $j_{_{\rm th}}$ is the spectral emissivity function ($\propto e^{-\e/kT} 
T^{-1/2}$) and $K_{\rm g}$ is a density dependent normalization factor.
      
\subsection{\bf Compact nucleus}

Evidence for a link between intense star formation and nuclear activity has 
grown steadily in recent years (e.g., Della Ceca et al. 2001; see 
Veilleux 2000 for a recent review). The apparent correlation -- deduced for 
nearby galaxies -- between the mass of the (dormant) nuclear BH and the mass 
of the spheroid suggests a direct link between the formation of spheroids 
and the growth of central BHs. Since a SB is a natural consequence of the 
dissipative gaseous processes associated with spheroid formation, an early 
(high-$z$) SB/AGN connection is implied by these results. The presence of 
circumnuclear SBs in many local AGN also suggests a local ($z \sim 0$) SB/AGN 
connection. This is important since a large contribution from hidden AGN 
would change the star formation history of the universe as deduced from 
galaxy luminosity functions, as well as our views on the history of the 
cosmic chemical enrichment and of the importance of feedback processes in 
the early universe. Moreover, the contribution of heavily-obscured 
AGN to the IR and the X-ray background should also be be accounted for. 

The fueling of AGN requires mass accretion at a rate
$$
\dot M ~ \sim ~ 1.7 \, \bigl( {\eta \over 0.1} \bigr)^{-0.1} \, \bigl( {L 
\over 10^{46} ~{\rm erg ~ yr}^{-1} } \bigr) M_\odot ~{\rm yr}^{-1} \,,
\eqno(17)
$$
where $\eta$ is the mass-to-energy conversion efficiency. For low-luminosity 
AGN (LLAGN) (e.g., Seyfert galaxies), eq.(19) implies a modest accretion 
rate, $\dot M \sim 0.01 \, M_\odot \,{\rm yr}^{-1}$, to power the activity 
level over timescales $\sim 10^8$ yr; a small fraction of the total gas 
content of typical host galaxy is then sufficient to fuel LLAGN. A broad 
range of mechanisms, including intrinsic processes (e.g., stellar 
collisions and winds, dynamical friction of giant molecular clouds against 
stars; nuclear bars or spirals produced by gravitational instabilities in 
the disk) and external processes (e.g., "minor" galaxy interactions and 
mergers), may be at work in LLAGN. Observationally, the former seem to be 
favored over the latter: e.g., ejecta from a nuclear star cluster may be 
enough for Seyferts and other LLAGN to keep their nuclei active (e.g., 
Maeda et al. 2001).

For high-luminosity AGN (HLAGN), meeting the stringent requirement on 
$\dot M$ implied by eq.(17) very likely requires external processes, such 
as major galaxy interactions or mergers, to be involved in triggering and 
sustaining the high activity level over $\sim 10^8$ yr. In fact, {\it (a)} 
substantial evidence exists that at least some HLAGN result from gas-rich 
mergers; {\it (b)} classical double (FR II) radio galaxies show tidal tails 
and other signs of interaction; {\it (c)} evidence for recent or ongoing 
galactic interactions is seen in several QSOs; {\it (d)} many radio galaxies 
and QSOs show the presence of abundant molecular gas (an essential ingredient 
of star formation), or the spectroscopic signature of recent star formation; 
{\it (e)} the FIR excess observed in several radio galaxies and QSOs is 
attributable to star formation. Ultra-luminous IR galaxies (ULIRGs) may 
represent the clearest observational link between galaxy mergers, SBs and 
powerful AGN: {\it i)} nearly all of them show strong signs of advanced tidal 
interactions; {\it ii)} all are very rich in molecular gas within the 
innermost kpc of the galaxy; {\it iii)} there is a varied level of activity 
in their nuclei, including strong emission lines, characteristic of 
starbursting stellar populations, and -- in $\sim$30\% of cases -- broad or 
high-ionization emission lines that suggest the presence of a powerful AGN 
coexisting with the SB (e.g.: Franceschini et al. 2000, Keil et al. 2001). 
The fraction of AGN-dominated ULIRGs is significantly larger among objects 
with high IR luminosities and warm IR colors. The relative dominance of AGN 
or SBs in ULIRGs may depend on local and short-term conditions (e.g., 
compression of the circumnuclear ISM as a function of gas content and galaxy 
structure, local accretion rate onto the central BH), as well as the global 
state of the merger. Suggestions of the existence of a merger-induced 
sequence "SBs $\rightarrow$ cool ULIRGs $\rightarrow$ warm ULIRGs 
$\rightarrow$ QSOs" imply that SB ages should increase along the sequence. 
This prediction should be testable with detailed spectroscopic data. 

The moderate luminosities of local SBGs indicate that if a nuclear source 
is present it is of the LLAGN type. {\it ASCA} measurements suggest that 
spiral galaxies often host nuclear LLAGN with $L_{\rm 0.5-10 \,keV} \sim 
10^{40} - 10^{41}$ erg s$^{-1}$ (Ishisaki et al. 1996). {\it ROSAT}/HRI 
detections of short-time variability (Collura et al. 1994) support this view. 
In M82, {\it ASCA} hard X-ray data have shown a nuclear unresolved point 
structure with a long-term flux variability by a factor of $\mincir 4$, 
corresponding to a point source luminosity of $L_{\rm 2-10\, keV } 
\sim 10^{40}$ erg s$^{-1}$ (see Ptak \& Griffiths 1999; {\it RXTE} data 
also suggest temporal variability in M82, see Rephaeli \& Gruber 2001; 
{\it Chandra}/HRC observations have shown that the source lies $\sim$160 pc 
away from the dynamical center of M82, see Kaaret et al. 2001 and Matsumoto 
et al. 2001). Since the observed X-ray luminosity in M82 is a lower limit 
to its Eddington luminosity, the implied BH mass is $\magcir$500$\,M_\odot$ 
(Ptak \& Griffiths 1999; see also Strickland et al. 2001 and Dahlem et al. 
1995), unless the emission is beamed (King et al. 2001). (The lack of reported 
temporal variability in other spectral bands in M82 may be explained as due 
to the large absorption column, and the fairly complicated pattern of 
activity that would make it hard to detect the temporal signature of LLAGN 
at other wavelengths.) Comparison of X-ray and FIR properties permits to 
disentangle SB and LLAGN emissions within the sources (Levenson et al. 
2001a,b).

Possible LLAGN contribution to the spectrum of SBGs can be represented as:
$$
f_{_{\rm LLAGN}}(\e) ~\propto~ \e ^{-\alpha}
\eqno(18)
$$
with a photon index $\alpha \sim 1.6$. A more accurate representation would 
include: {\it i)} a reflection component, in the form of a hardening of the 
spectrum at $\magcir$10 keV (i.e., the onset of a broad bump extending from 
$\sim$10 keV to $\sim$60$-$100 keV, that arises from the downscattering of 
the energetic photons of the primary spectrum by the optically thick warm 
thermal matter of the accretion disk; cf. Lightman \& White 1988, Magdziarz 
\& Zdziarski 1994); and {\it ii)} a Comptonization effect in the form of a 
downward bending at $\sim$60$-$300 keV, arising from the interaction of 
radiation with hot embedding ionized plasma (cf. Sunyaev \& Titarchuk 1980). 
But, given the substantial uncertainty in assessing the significance of 
AGN-like emission in SBGs, we will not consider a nuclear contribution 
further in this paper. 

\section{The composite X-ray spectrum of a SBG}

Having discussed the most relevant emission processes that occur in a SB 
	\footnote{
	GRBs ($\gamma$-ray bursts), short and intense bursts of 0.1-1 
	Mev photons (Piran 1999) followed by lower-frequency afterglows (van 
	Paradijs et al. 2000), do not contribute to the mean SBG emission. 
      Based on energy and variability arguments, GRBs have been related to 
	the final stages of supermassive ($35 \mincir M/M_\odot \mincir 50$) 
	stars, where non-spherical core collapse would produce beamed 
	$\gamma$-ray emission (Mac Fadyen \& Woosley 1999). In this case: 
	{\it i)} for a standard IMF slope, candidate GRB progenitors are 
	$\sim 10^{-2}$ of SN progenitors; {\it ii)} beaming effects introduce 
	an extra factor $(1- {\rm cos}\theta)/2 \sim 10^{-3}$ (with an opening 
	angle $\theta \sim 5^{\circ}$, see Frail et al. 2001); {\it iii)} for 
	a Galactic SN rate of $\sim 10^{-2}$ yr$^{-1}$, the predicted GRB rate 
	is then $\sim 10^{-7}$ yr$^{-1}$. [A consequence of this: since 
	there are ${\cal O}(9)$ galaxies of luminosity $\sim L_B^{\star}$ within 
	the horizon, one expects to observe ${\cal O}(2)$ GRBs yr$^{-1}$, in 
	agreement with {\it BATSE} results (Paciesas et al. 1999).] As the 
	X-ray phase of a GRB afterglow lasts $\sim 10^{-2}$ yr, we expect 
	$\sim 10^{-9}$ GRB X-ray afterglows per galaxy. The contribution of 
	GRBs to the SBG emission is then negligible.}
we now construct the synthetic (mean) X-ray spectrum of SBGs. 
Its shape is given by 
$$
f(\e) ~=
$$
$$
=~ {
[N \bar L_{\rm x} \bar J(\e)]_{_{\rm H}}  \, + \,
[N \bar L_{\rm x} \bar J(\e)]_{_{\rm L}}  \, + \,
[N \bar L_{\rm x} \bar J(\e)]_{_{\rm SNR}}  
\over 
[N \bar L_{\rm x}]_{_{\rm H}}  \, + \,
[N \bar L_{\rm x}]_{_{\rm L}}  \, + \,
[N \bar L_{\rm x}]_{_{\rm SNR}}  
 } ~+
$$
$$
+~ f_{\rm C}(\e) \, +\, f_{\rm g}(\e)   \,.
\eqno(19)
$$

In our framework the combined "stellar" contribution [first RHS term of 
eq.(19)] is taken to have no degrees of freedom -- the fractional number 
densities of the various classes of stellar sources are fixed by the 
synthetic stellar population model, and the respective observed spectra. 
The degrees of freedom in modelling the diffuse component are: (1) the 
gas/stars mass ratio (reflected in the parameter $K_{\rm g}$); (2) the 
chemical abundance of the gas; and (3) the strength of the FIR radiation 
field (for a given energy distribution of the electrons, as deduced from 
the observed radio spectrum), reflected in the strength $K_{_{\rm C}}$ of 
the Compton emission.

In Fig.1 we show the resulting spectrum for the "standard" set of model 
parameters (as specified in eqs.5, 9, 15, 16, 17). 

\begin{figure}
\vspace{6.0cm}
\includegraphics{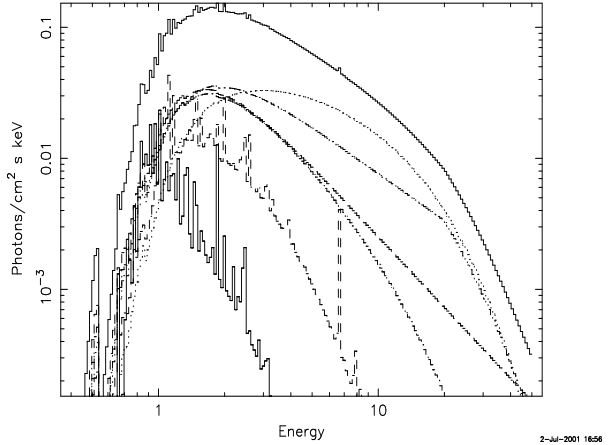}
\caption{The normalized template spectrum of SBGs. In increasing flux at 
6 keV, the various components are: galactic wind, SNRs, faint LMXBs, 
Compton emission, HMXBs (no Fe-K emission line at 6.7 keV included), 
and bright LMXBs. The assumed luminosities (in the 0.5-50 keV band) are: 
log$L_{\rm x} = 37.7$ for HMXBs and bright LMXBs, 37.0 for SNRs, and 36.7 
for faint LMXBs. The galactic-wind and Comptonized emissions are scaled to 
0.1 of the total flux at 1 keV and 10 keV, respectively. The spectral 
components are normalized in photon flux in the 0.5-50 keV band. The spectrum 
is absorbed through a hydrogen column density of $n_{\rm H}= 10^{22}$ 
cm$^{-2}$.}
\end{figure}

\section{Observed SBG spectra: M82 \& N253}

The power of X-ray spectral analysis as a diagnostic tool of SBGs is mainly 
expressed in the ability to identify the leading components by means of the 
presence of distinctive spectral features. In order to assess the relative 
contributions of the various processes identified in the previous section, 
we focus on the spectral insights that can be learned in the best studied 
nearby SBGs, M82 and N253. Before doing that, we should comment on some 
implications of the results obtained in sections 2 and 3.

\noindent
$\clubsuit$ At high energies ($\e \magcir 10$ keV) the dominant contributions 
are from bright LMXBs and -- possibly -- nonthermal emission. LMXBs 
constitute $\sim 80\%$ of the total population of bright (log$L_{\rm x} > 
37$) X-ray binaries and their population-averaged spectrum is fairly flat. 
Nonthermal emission results from either {\it (a)} Compton scattering of 
SN-accelerated relativistic electrons off the FIR and CMB fields, and/or 
{\it (b)} activity from a compact nuclear source. In M82, there is some 
evidence for the latter emission, as is possibly indicated by appreciable 
temporal variability seen in {\it RXTE} data (Rephaeli \& Gruber 2001). 
This is in accord with earlier {\it ASCA} results, in which an unresolved 
point-like source -- possibly an obscured low-luminosity AGN -- was 
determined to dominate the 2-10 keV flux, and exhibited spectral variability 
on a characteristic $\sim$5 year period (Tsuru et al. 1997). 

\noindent
$\clubsuit$ At intermediate energies ($2 \mincir \e/{\rm keV} \mincir 10$) 
bright LMXBs are the stellar component that dominates the continuum (by 
virtue of their abundance, luminosity, and spectral hardness), with possibly 
appreciable contribution also from Compton scattering. HMXBs, though 
constituting only $\sim 20\%$ by number of the bright binary population, 
may contribute crucially to this spectral range: Their relatively hard 
spectrum shows a pronounced Fe-K emission between 6.4 and 6.7 keV, with 
$EW_{\rm Fe-K} \sim 0.2-0.6$ keV (White et al. 1983). This feature has 
important implications on the chemical abundances as deduced from spectral 
analyses. In fact, recall that the "hard" component required to fit the {\it 
ASCA} and {\it BeppoSAX} data has been interpreted either as thermal ($kT 
\magcir $6 keV), as power-law ($\Gamma \sim 1.7$), or as combination of 
both. In the first case, the measured $EW_{\rm Fe-K}$ implies a low chemical 
abundance ($\mincir$0.3 solar) in the hot gas, while if the thermal and 
nonthermal contributions are comparable, then the deduced chemical abundance 
could be roughly solar (Ptak et al. 1997, Cappi et al. 1999, Zezas et al. 
1998, Della Ceca et al. 1999, Griffiths et al. 2000). Now, an abundance 
$Z \sim 0.3 Z_\odot$ is probably uncomfortably low for a medium that has been 
stirred up and enriched by SN activity. In addition, the very presence of 
large amounts of $\magcir$5 keV gas poses substantial energetics and 
hydrostatic equilibrium problems.
      \footnote{Very hot ($\magcir$5 keV) thermal emission may come from 
      localized regions, such as the inner SB region, where the highest 
      concentration of SN events occur. In fact, temperatures $kT \sim 5-8$ 
      keV can be momentarily reached in the very short Phase I of the 
      evolution of SNRs, lasting ${\cal O} (2)$ yr (e.g., Charles \& Seward 
      1995; see Pietsch et al. 2001 for N253).} 
If the "hard" component were (mainly) due to the emission from X-ray binaries 
and possibly nonthermal Compton emission, the consequences would be that 
$EW_{\rm Fe-K} \mincir 0.4$ (i.e., emission from HMXBs will be overshadowed 
by other emission) and there would be no need for a high-$kT$, low-$Z$ plasma. 

\noindent
$\clubsuit$ At energies $\e \mincir 2$ keV, the low temperature ($kT \leq 1$ 
keV) diffuse plasma resulting from the interaction between the hot, 
low-density galactic winds and the cold, high-density ISM is an important 
component.

\noindent
$\clubsuit$ The contribution of young, ejecta-dominated SNRs to the continuum 
X-ray emission of a SBG is probably minor. Indeed, there seems to be 
little overlap between the X-ray source position in M82 
(based on high-resolution {\it Chandra} measurements), and the positions of 
radio-detected young (i.e., brighter) SNRs (Griffiths et al. 2000). Thus, 
interpretation of X-ray data based on the assumption that the $\magcir$2 keV 
emission comes mostly from SNRs (e.g., Pietsch et al. 2001) is doubtful. On 
the other hand, emission from older, cooler ($\sim$1 keV) remnants might be 
hard to distinguish from the diffuse thermal emission, similarly 
characterized by $\sim$1 keV. Remarkably, SNRs might provide a significant 
contribution to the observed 6.7 keV Fe-K line (cf. Fig.1), further weakening 
the case for a thermal interpretation of the 2-10 keV "hard" component. 
Some caution, however, is in order. It should be emphasized that the 
properties of SBG SNRs may differ from those of Galactic SNRs: owing to the 
variety of environment (e.g., the ISM is much denser in SBGs) the ejecta will 
generally experience diverse (and complex) evolution patterns, which will in 
turn lead to a broad range of emission properties
      \footnote{This topic is relevant for evolutionary models of the 
      radio emission produced in compact SBs (P\'erez-Olea \& Colina 1995), as 
      well as for studies of alternative physical mechanisms (i.e., 
      mechanisms not requiring the presence of a supermassive-BH) that would 
      allow an explanation of the observed properties of AGNs (Terlevich et 
      al. 1992 and references therein).
(e.g., Cid-Fernandes et al. 1996). Consequently, even among SBGs the 
properties of SNRs may have a large scatter. It has also been suggested that 
SNRs can contaminate the bright end of the X-ray--binary luminosity function, 
and hence the integrated spectra of the host galaxies (Wu et al. 2001). }

\begin{figure}
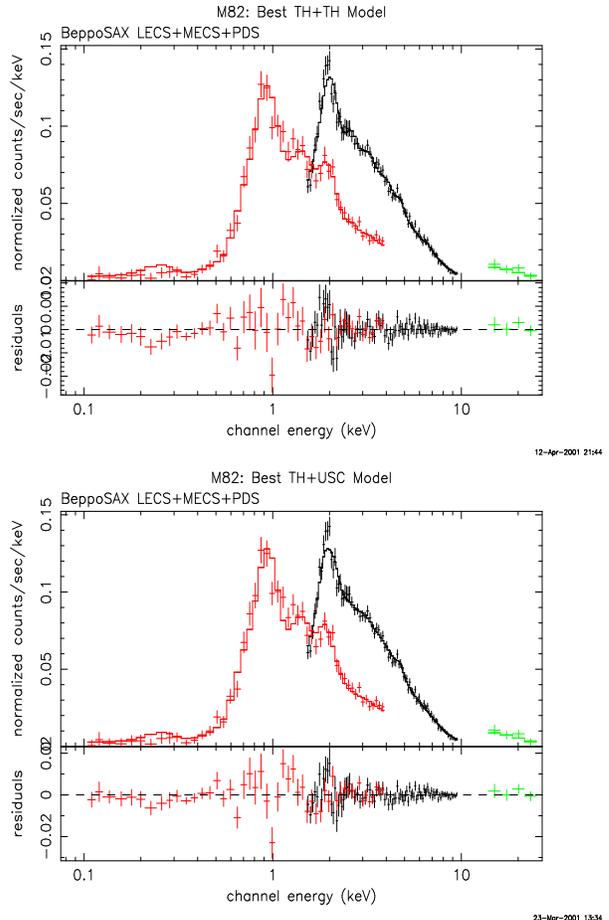

\vspace{6.0cm}
\includegraphics{MS1760f2a.cps}
\vspace{.2cm} 
\vspace{6.0cm}
\includegraphics{MS1760f2b.cps}
\caption{{\it BeppoSAX} spectrum of M82 with superposed: {\it (top)} the 
double-thermal model (Cappi et al. 1999); {\it (bottom)} the thermal+USC model.
The residuals are shown in the lower sections of the panels. Red, black, and 
green data points denote the LECS, MECS and PDS data, respectively.}
\end{figure}

\begin{figure}
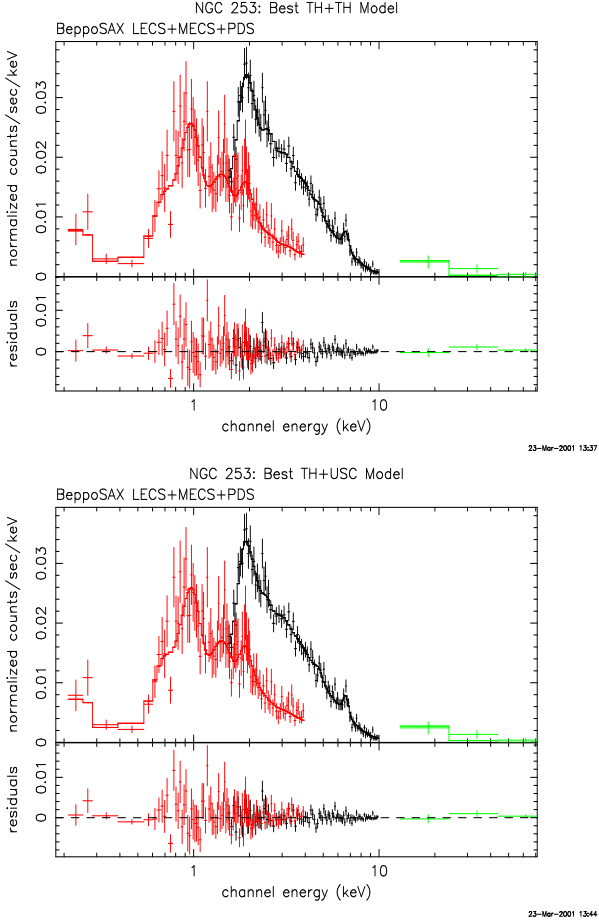

\vspace{6.0cm}
\includegraphics{MS1760f3a.cps}
\vspace{6.0cm}
\vspace{.2cm} 
\includegraphics{MS1760f3b.cps}
\caption{{\it BeppoSAX} spectrum of N253 with superposed: ({\it top}) the 
double-thermal model (Cappi et al. 1999); ({\it bottom}) the thermal+USC 
model (a Gaussian fit the Fe-K emission complex). The residuals are shown 
in the lower sections of the panels. Red, black, and green data points denote 
the LECS, MECS and PDS data, respectively.}     
\end{figure}

Based on the above summary of sources of X-ray emission in SBGs, we are led 
to the consideration of binary systems as the main contributors to the 2-10 
keV emission of SBGs, with the possibility of appreciable contribution also 
from Compton scattering. To check this hypothesis, we have re-analyzed the 
0.5-20 keV {\it BeppoSAX} LECS+MECS+PDS spectra of M82 and N253 (see Cappi 
et al. 1999). For M82, the best-fitting model ($\chi_\nu^2=1.24$ with 
DOF=138) of Cappi et al. includes two thermal components, one with $kT=0.70$ 
keV and the other (self-absorbed through $n_{\rm H}= 0.58 \times 10^{22}$ 
cm$^{-2}$) with $kT=8.20$ keV, both chemically unevolved ($Z << Z_\odot$). 
The model for N253 is similar ($\chi_\nu^2=1.06$ for DOF=184): the two 
thermal components have $kT=0.81$ keV and 5.75 keV, respectively (with the 
latter component self-absorbed through $n_{\rm H}= 1.19 \times 10^{22}$ 
cm$^{-2}$), and both chemically unevolved ($Z \mincir 0.3 \, Z_\odot$). 
(See Cappi et al. 1999 for details on both models.) As a check of the 
hypothesis that in both galaxies the hard component is mainly due to 
massive-binary emission, we replaced the harder component in these models 
with an $\e^{-\Gamma} e^{-\e/kT}$ component (i.e., the USC profile used to 
describe the spectra of LMXBs, the more abundant population of X-ray 
binaries), plus a Gaussian to fit the Fe-K emission complex (as needed), 
and found the best-fit parameters with the soft thermal component kept 
frozen. The resulting parameters of the USC component are: $\Gamma=1.4 
\pm 0.1$ and $kT=9.2 \pm 2.0$ keV (self-absorption: $n_{\rm H}= 0.72 
\times 10^{22}$ cm$^{-2}$) for M82; and $\Gamma=1.55 \pm 0.25$ and $kT=7.3 
\pm 2.7$ keV (self-absorption: $n_{\rm H}= 1.5 \times 10^{22}$ cm$^{-2}$), 
with the Fe-K emission feature centered at $E=6.7$ keV, and with $EW_{\rm 
Fe-K} =0.34$ keV for N253. (Note that the PDS data points have negligible 
effects on our results.) The resulting values of $\Gamma$ and $kT$ are well 
within the ranges observed in LMXB spectra. For both M82 and N253, then, the 
thermal+USC fits are as satisfactory as the original double thermal models 
(see Figs.2,3). These results are robust. In fact, the best-fit thermal+USC 
models:

\noindent
$\bullet$ {\it M82}. 
Thermal component: $kT=0.74 \pm 0.03$ keV (with $Z \sim Z_\odot$); 
USC component: $\Gamma=1.39 \pm 0.14$ and $kT=8.7 \pm 2.0$ keV 
(self-absorption: $n_{\rm H}= 0.83 \times 10^{22}$ cm$^{-2}$); 
goodness of fit: $\chi_\nu^2=1.22$ (DOF=138);

\noindent
$\bullet$ {\it N253}. 
Thermal component: $kT=0.83 \pm 0.06$ keV (with $Z \sim Z_\odot$); 
USC component: $\Gamma=1.6 \pm 0.4$ and $kT=7.9 \pm 4.1$ keV 
(self-absorption: $n_{\rm H}= 1.56 \times 10^{22}$ cm$^{-2}$); 
Fe-K emission feature at fitted with a Gaussian): $E=6.7$ keV, $EW_{\rm 
Fe-K}=0.4$ keV; goodness of fit: $\chi_\nu^2=1.05$ (DOF=179). 

\begin{figure}
\vspace{6.0cm}
\includegraphics{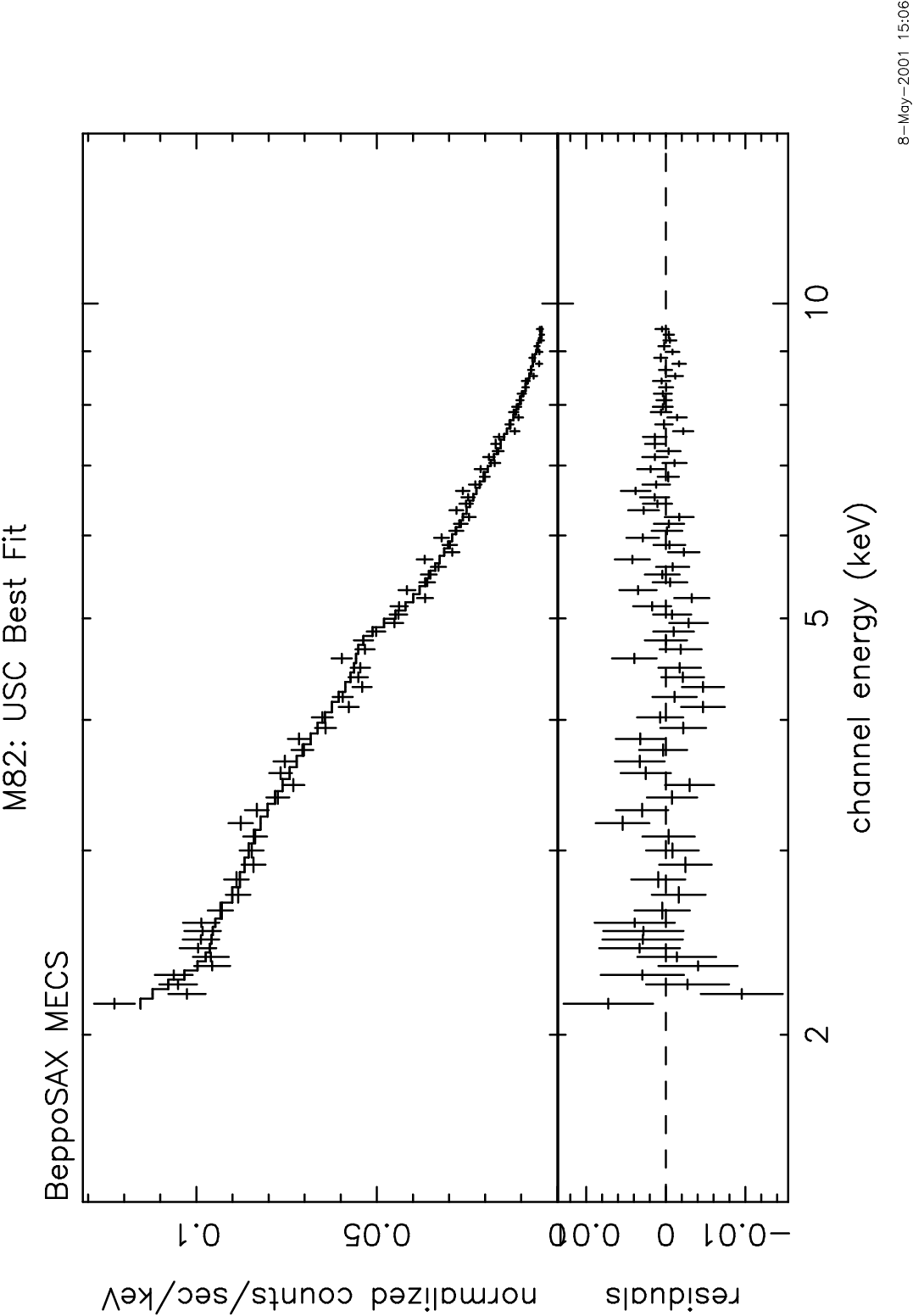}
\vspace{6.0cm}
\includegraphics{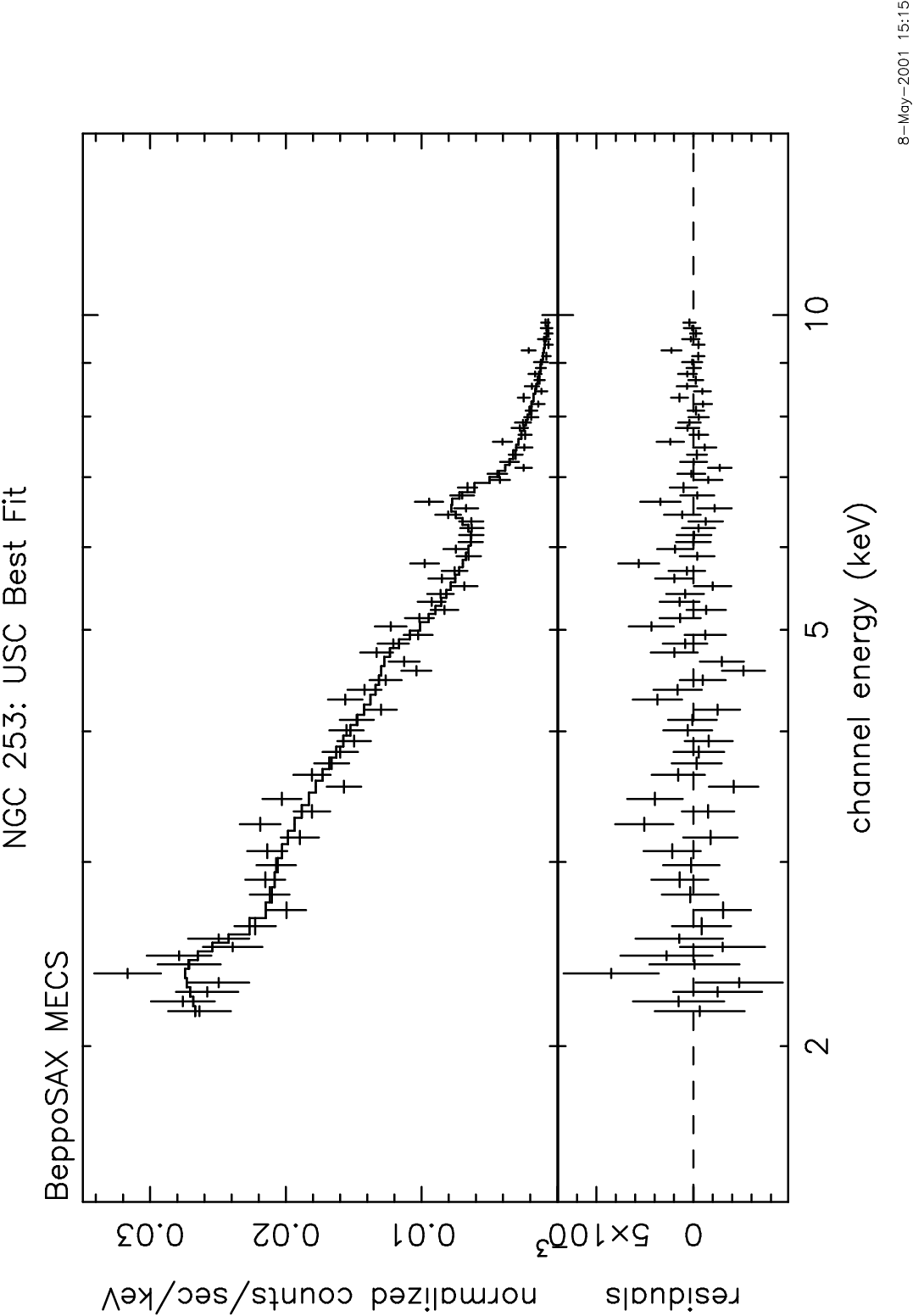}
\caption{
The 2-10 keV {\it BeppoSAX}/MECS spectra of M82 ({\it top}) and N253 ({\it 
bottom}) with superposed best-fitting USC models (M82: $\Gamma = 1.2$, 
$kT=6.75$, $n_{\rm H}= 0.44 \times 10^{22}$ cm$^{-2}$; N253: $\Gamma = 1.59$, 
$kT=7.72$, $n_{\rm H}= 1.07 \times 10^{22}$ cm$^{-2}$). The emission features 
[S{\tt XV} at $\sim$2.4 keV, and (for N253) Fe-K at 6.7 keV] are fitted with 
a Gaussian. The differences in the values of $\Gamma$ between the present 
fits and those shown in Fig.3 are mainly due to the additional separate 
fitting -- present here but not in Fig.3 -- of the S{\tt XV} line at $\sim$2.4 
keV. The residuals are shown in the lower sections of the panels. }
\end{figure}

\noindent
Furthermore, fitting the 2-10 keV spectra of M82 and N253 with only the USC 
model gives good fits, with the values of the USC parameters remaining 
virtually unchanged (see Fig.4). 

As previously mentioned, the stellar component of the X-ray continuum has no 
degrees of freedom once the stellar population model has been selected, and 
the relevant spectral properties have been observed. In our treatment, with 
the Iben et al (1995a,b) Galactic synthetic model and a bright-to-faint LMXB 
luminosity ratio of 10, the predicted total "stellar" spectrum is fairly flat 
      \footnote{Changing the bright-to-faint LMXB luminosity ratio does not 
      significantly alter the results. Lowering the assumed bright-to-faint 
      LMXB luminosity ratio and steepening (within observational errors) the 
      mean HMXB causes only a slight steepening of the predicted "stellar" 
      spectrum. Thus, the flatness of the predicted "stellar" spectrum is a 
      relatively secure result in our treatment.} 
in the 2-10 keV band ($\Gamma \mincir 1$; see Fig.5), because in this region 
the spectrum is dominated by the flat-spectrum bright LMXBs and HMXBs, while 
the data require $\Gamma \magcir 1.2$. Therefore, either {\it (a)} most of 
the massive X-ray binaries responsible for the hard component have a 
relatively steep spectrum, $\Gamma \magcir 1.4$ (and hence resemble faint 
Galactic LMXBs), or {\it (b)} there is also an appreciable contribution due 
to Compton scattering (with $\Gamma > 1.5$), resulting in a suitably steep 
spectrum (see Fig.6). In the former case, if the spectral properties of X-ray 
binaries are the same in SBGs as in the Galaxy, the (fractional) population 
of low-luminosity LMXBs should be higher in SBGs than in the Galaxy in order 
to give a suitably steep integrated "stellar" spectrum. Consequently in SBGs 
the form of the binary birthrate function should differ from the Galactic 
one given in eq.(2), (e.g.) by a factor $q^{-\delta} a^{\eta}$ with $\delta 
\simeq 1/2$ and $\eta \simeq 3/4$, so as to imply a higher-than-Galactic 
fraction of low-luminosity LMXBs and hence a steeper composite spectrum 
($\Gamma \sim 1.4$) as required observationally. 

\begin{figure}

\vspace{3.4cm}
\includegraphics{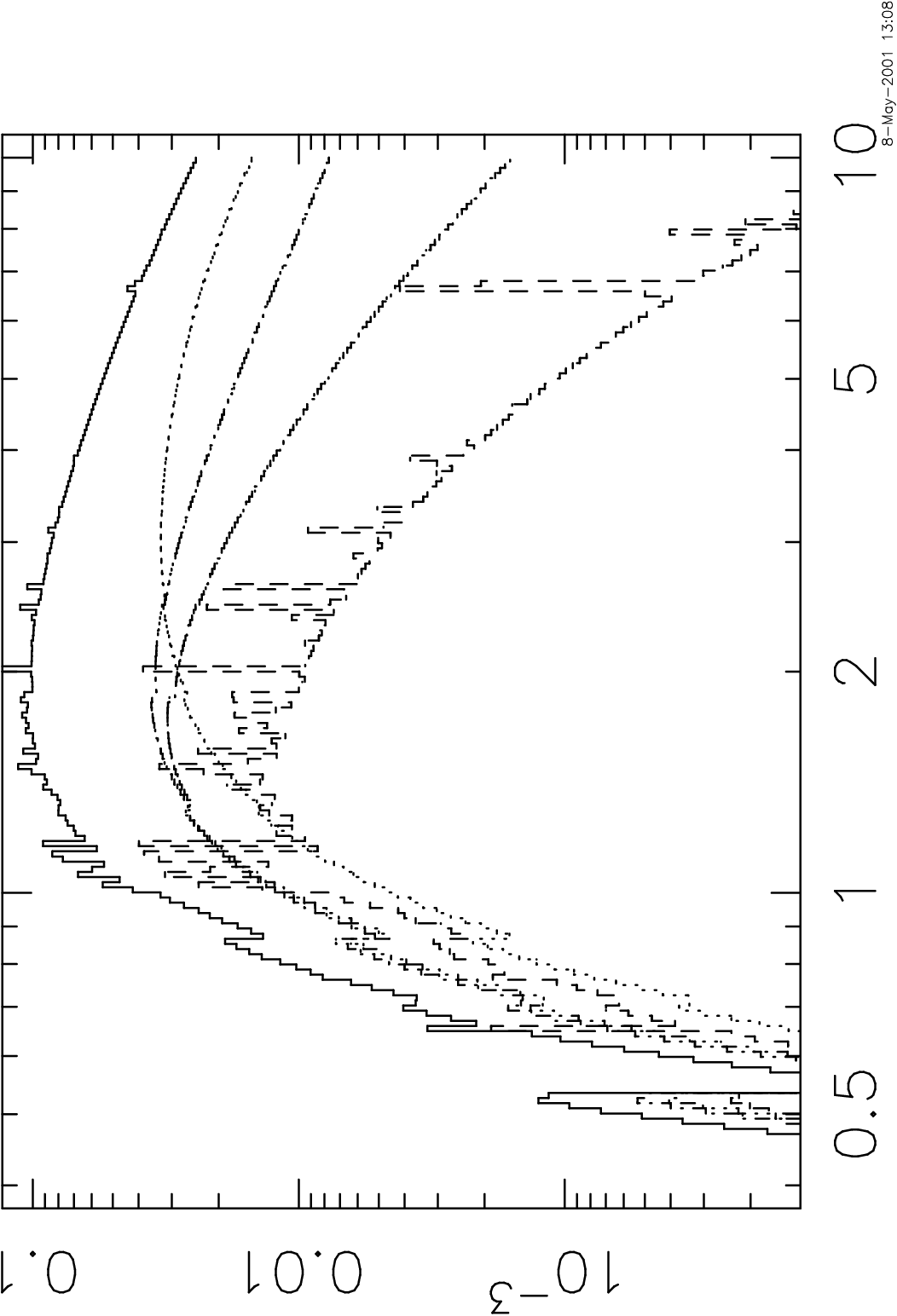}
\hspace{.5cm} 
\includegraphics{MS1760f5b.ps}
\vspace{.2cm} 

\vspace{3.4cm}
\includegraphics{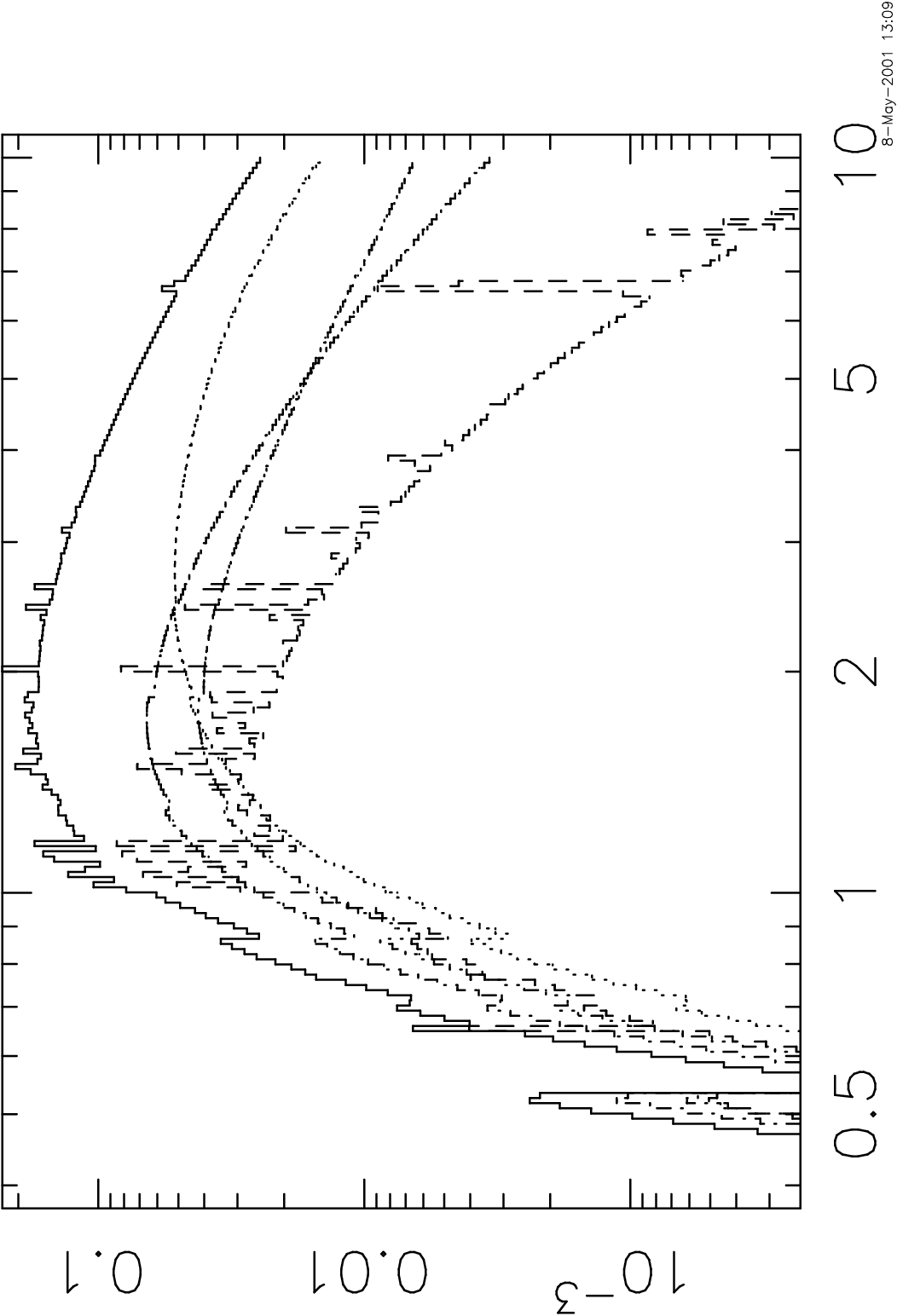}
\hspace{.5cm} 
\includegraphics{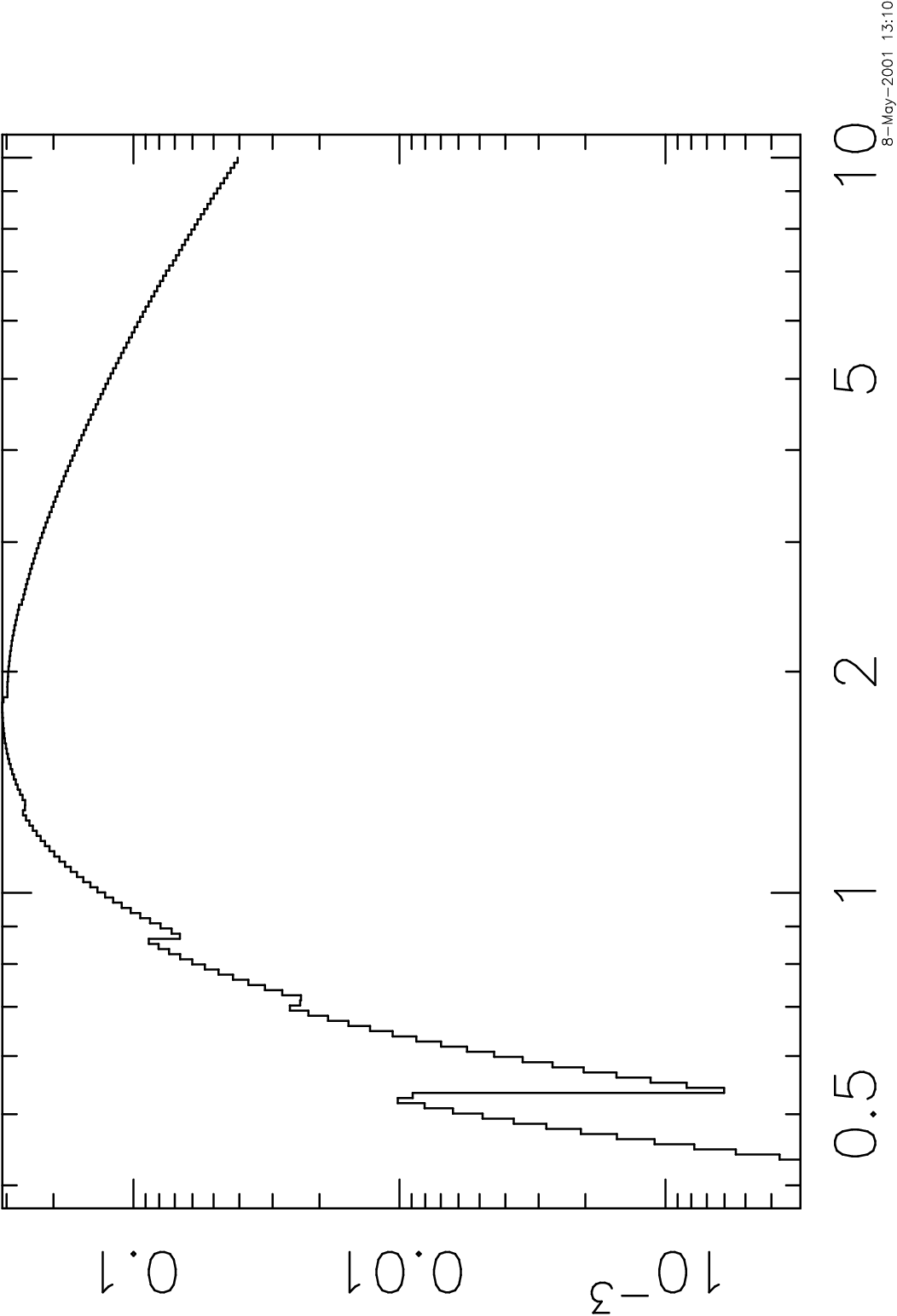}

\caption{The integrated "stellar" emission for two different sets of 
parameters: "standard" (see Fig.1; {\it top left}), and with the HMBX and 
bright-LMXB emissions steepened [corresponding, respectively, to $\gamma=1.4$ 
and to $kT=5$ keV ({\it bottom left})]. The "stellar" spectra show, in 
ascending order at 7 keV: SNRs, faint LMXBs, HMXBs (no Fe-K emission 
line at 6.7 keV included), and bright LMXBs, with luminosities and 
normalizations as specified in Fig.1. The spectra are absorbed through 
a hydrogen column density of $n_{\rm H}= 10^{22}$ cm$^{-2}$. The USC 
profiles have: $\Gamma = 0.8$, $kT=12$ keV, $n_{\rm H}= 10^{22}$ cm$^{-2}$ 
({\it top right}); and $\Gamma = 1.0$, $kT=11$ keV, $n_{\rm H}= 0.8 \times 
10^{22}$ cm$^{-2}$ ({\it bottom right}).}
\end{figure}

Another possibility involves a more radical variation from our assumed 
theoretical scenario. In a SB the formation of low-mass stars could be 
suppressed if SN blast waves of more rapidly forming massive stars did 
disrupt the slowly forming less massive stars before these have completed or 
even have reached their Hayashi tracks. The ensuing stellar initial mass 
function (IMF) 
would be truncated so that only stars with mass above some low-mass cutoff 
will form. Indeed, evolutionary models of the SB in M82 suggest that the IMF 
is likely to be restricted to $M \magcir 3\, M_\odot$ (Doane \& Mathews 1993; 
Rieke et al. 1993)
      \footnote{Note that also the putative SB at the Galactic center is 
      best modelled with a truncated IMF, with $M>10\,M_\odot$ (Lipunov et 
      al. 1996a).}.
In this case no LMXBs would form, and the binary population of the SB would 
only consist of HMXBs. The resulting 2-10 keV synthetic spectrum would then 
be dominated by HMXBs, and hence the USC fit would have $\Gamma \rightarrow 
\gamma = 1.2^{+0.2}_{-0.2}$ [cf. eq.(5); see Fig.7]. Note that M82 does have 
$\Gamma \simeq 1.2$ (see Fig.4). As a matter of fact, even if the IMF did 
retain its universal properties of shape and mass domain so that the formation 
of low-mass stars were not inhibited, a galaxy undergoing one isolated SB 
episode would not experience a SB-driven increase of the population of LMXBs. 
In fact, the time required for the $\mincir$1$\,M_\odot$ optical companion 
in a LMXB system to evolve out of the main sequence and come into Roche-lobe 
contact (and hence start the X-ray bright phase) by far exceeds a typical 
SB lifetime. So, during one isolated SB episode there is time for only HMXBs 
to form: HMXBs would then be the only type of X-ray binaries contributing to 
the X-ray emission from an isolated SB. (Of course, the LMXBs associated with 
the background old stellar population of the galaxy would also contribute to 
the overall emission.) LMXBs could be important contributors to the X-ray 
emission of SBGs that have been undergoing recurrent bursts of star formation: 
such could be the case for, e.g., galaxies that are members of pairs with 
highly eccentric orbits, or galaxies that are found in crowded environments 
(e.g., compact groups: see Hickson et al. 1989): in both cases, tidal 
interactions would be recurrent -- whether periodically or aperiodically. 
Therefore, determining the type of X-ray binaries whose emission dominates 
the 2-10 keV luminosity of a SBG would provide a clue to understanding the 
star formation history of that galaxy.

\begin{figure}

\vspace{3.4cm}
\includegraphics{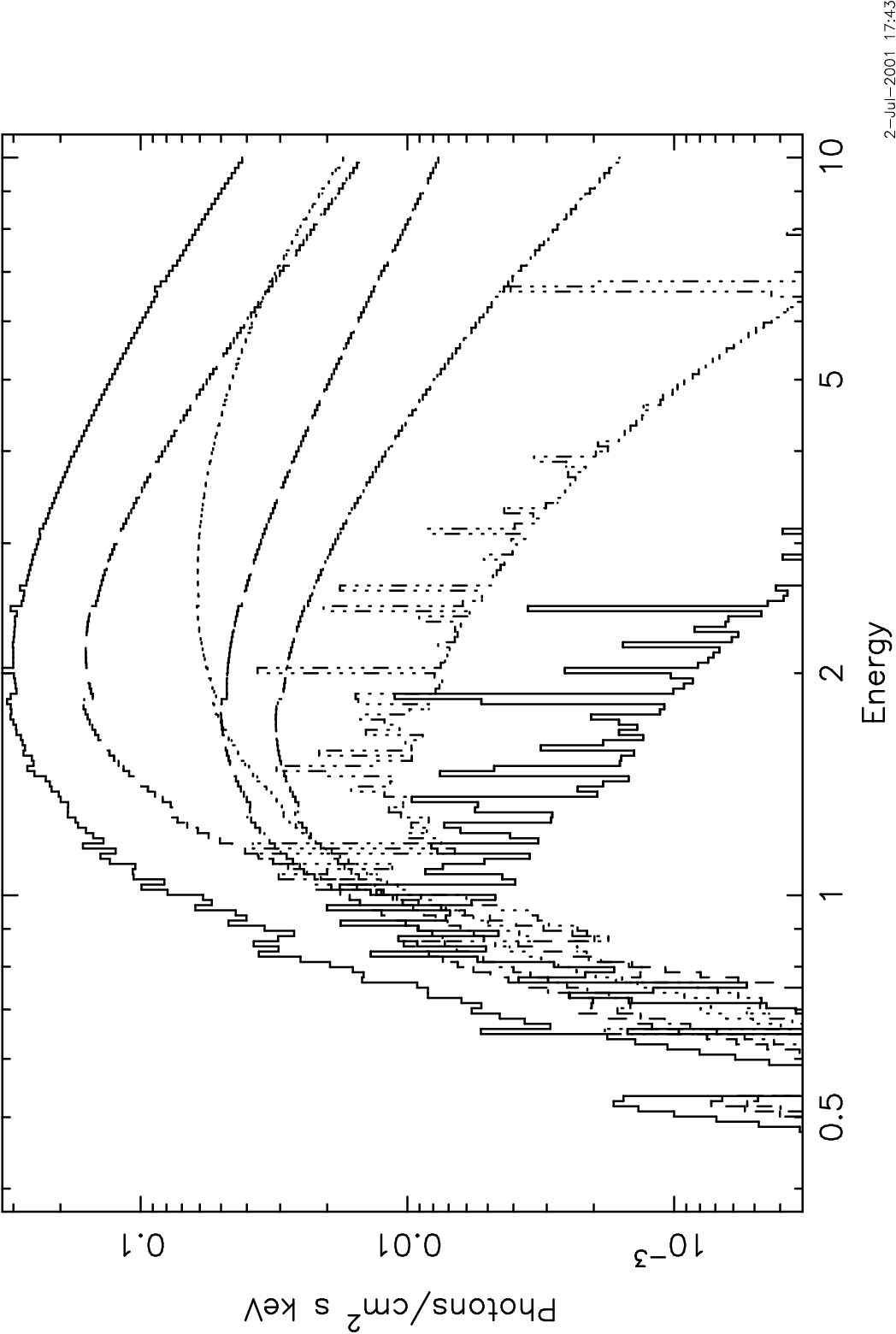}
\hspace{.5cm} 
\includegraphics{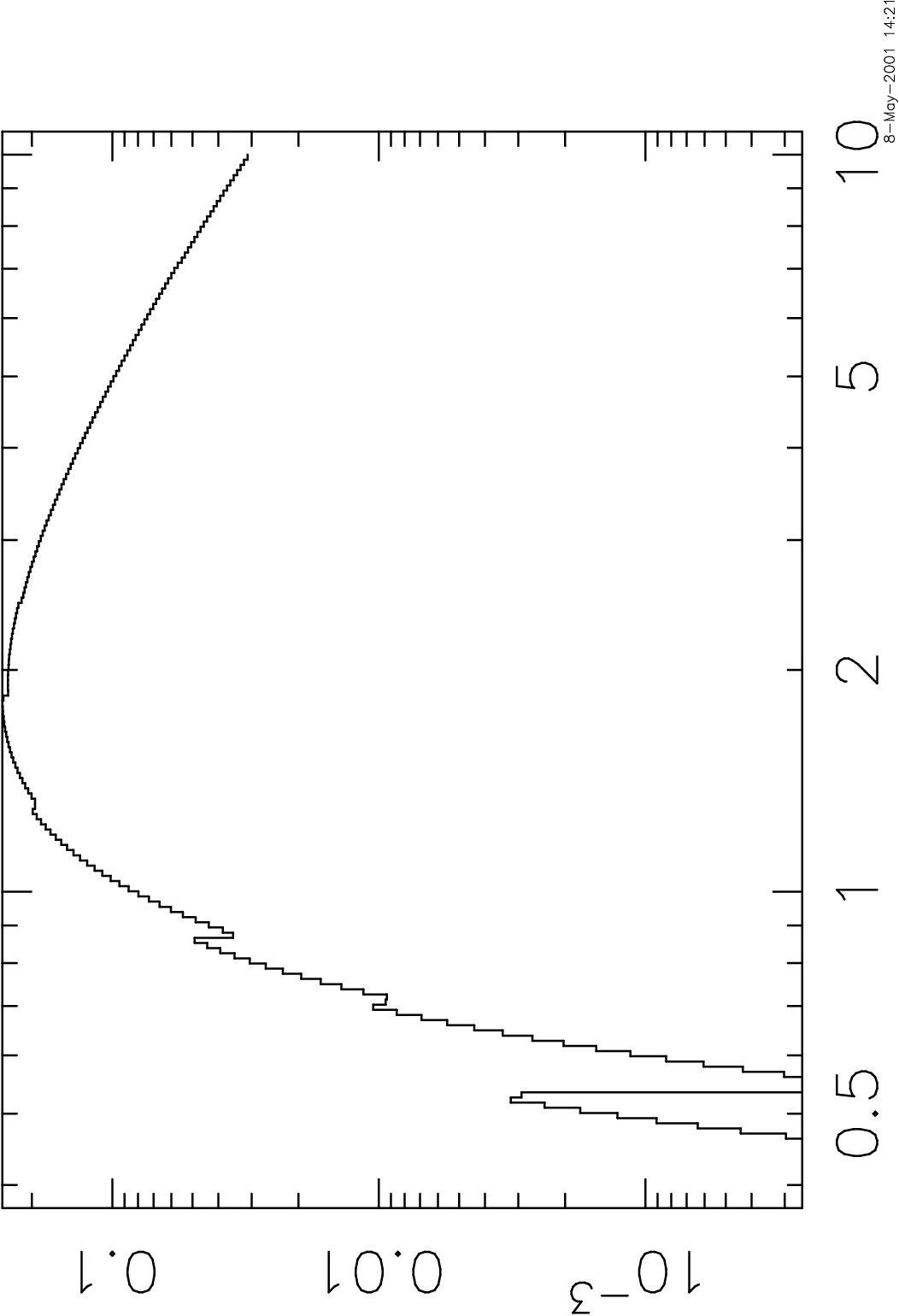}
\caption{A template SBG spectrum ({\it left}) vs. an USC profile with 
$\Gamma=1.2$, $kT=14$ keV ({\it right}), both with overall absorption 
through a hydrogen column density of $n_{\rm H}= 10^{22}$ cm$^{-2}$. The 
spectral components are (in ascending order at 3 keV): galactic winds, SNRs, 
faint LMXBs, HMXBs (no Fe-K emission line at 6.7 keV included), bright 
LMXBs, and Compton emission ($\beta=2$, self-absorbed through $n_{\rm H} = 
10^{22}$ cm$^{-2}$). The stellar emission is as in the bottom left case of 
Fig.5. The galactic-wind and Compton emissions are scaled to 0.1 of the total 
flux at 1 keV, and 0.35 of the total flux at 10 keV, respectively.}
\end{figure}

The arcsec angular resolution of current X-ray telescopes (e.g., {\it 
Chandra}) enables observational tests to discriminate between LMXBs and HMXBs 
as the main (stellar) contributors to the 2-10 keV emission in SBGs. For 
example, LMXBs contain low-mass stars and hence they should be partially 
associated with the bulge, while HMXBs (which contain, of course, more massive 
stars), should not (e.g., Pence et al. 2001): therefore LMXBs and HMXBs are 
expected to have different spatial distributions within a galaxy. Also, since 
a significant fraction of star formation occurs in clusters (e.g. M82: see 
O'Connell et al. 1995 in the optical, and Griffiths et al. 2000 in X-rays), 
and stellar-cluster ages can be inferred from optical data coupled with 
evolutionary models of HII clouds embedding a cluster (e.g., 
Garc{\'\i}a-Vargas et al. 1995a, 1995b), constraints can be put on the masses 
of the X-ray--emitting stars. Finally, as the spectral profile of the point 
sources is an efficient diagnostics of the nature of the accreting objects 
(Ostriker 1977; White \& Marshall 1984), source-by-source measurements of the 
X-ray spectral hardness could give us important clues for the identifications 
of such point-like sources
      \footnote{The need for individual spectral identifications arises 
      from the following argument. In the 2-10 keV band, HMXBs have a 
      power-law spectrum with photon index $\gamma= 1.2 \pm 0.2$ [see 
      eq.(5)]. In the same band, luminous ($>$10$^{37}$ erg s$^{-1}$) LMXBs 
      have a harder ($\propto e^{-\e/ (7.5\, {\rm keV}) }$) spectrum, while 
      less luminous ($<$10$^{37}$ erg s$^{-1}$) LMXBs have a softer ($\propto 
      \e^{-1.4} e^{-\e/(7.5\, {\rm keV})}$) spectrum [see eqs.(9a), (9b)]: 
      the resulting population-averaged spectrum [see eq.(11)] can be fitted 
      by a power law of photon index $\simeq$1.1 (see Fig.8). Therefore, the 
      effective spectral indexes of the integrated LMXB and HMXB spectra are 
      very similar, and compatible with each other. So, from their integrated 
      spectra the two populations of HMXBs and LMXBs can hardly be 
      distinguished from each other: hence the need for individual spectral 
      measurements of the point-like sources. }.

\begin{figure}

\vspace{3.4cm}
\includegraphics{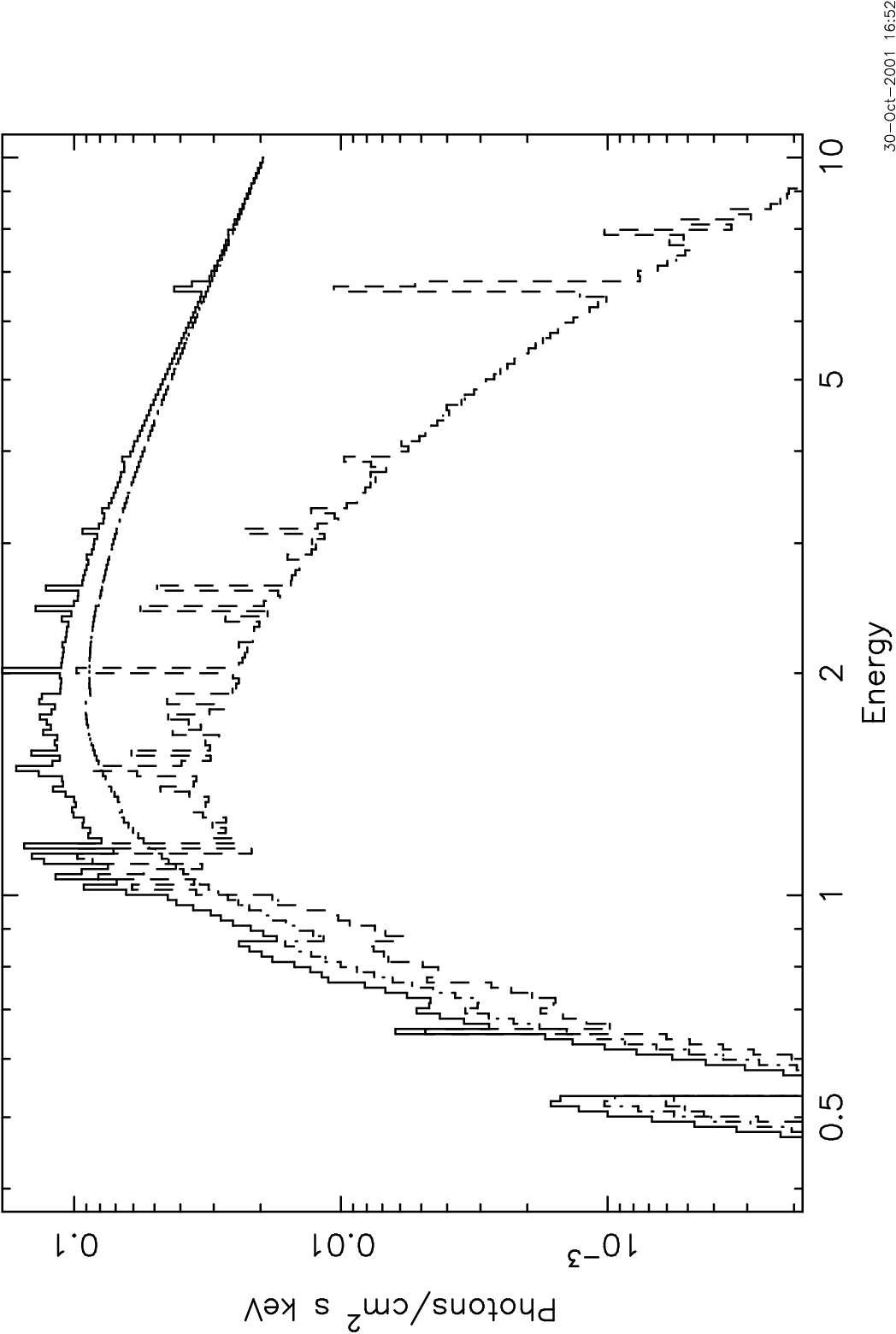}
\hspace{.5cm} 
\includegraphics{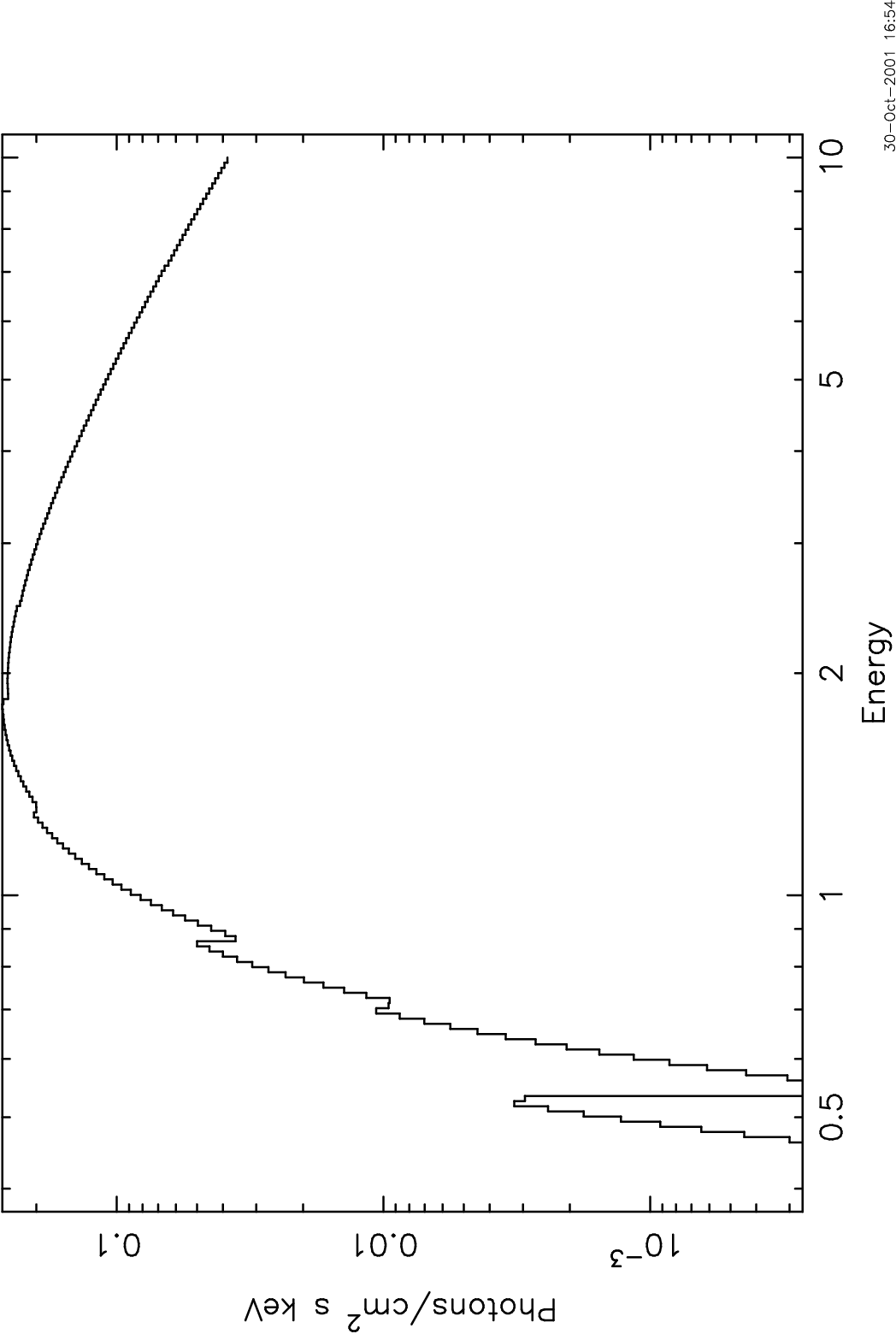}
\caption{
{\it Left}: an integrated SB "stellar" spectrum that includes (in ascending 
order at 3 keV) only SNRs and HMXBs (no Fe-K emission line at 6.7 keV 
included); the relevant parameters and normalizations are as in Fig.1. 
{\it Right}: an USC profile with $\Gamma=1.2$ and $kT=20$ keV. Both the 
"stellar" spectrum and the USC model are absorbed through a hydrogen column 
density of $n_{\rm H}= 10^{22}$ cm$^{-2}$. The spectrum simulates the X-ray 
emission from a stellar population having no LMXBs, such as it may occur in 
an isolated burst of star formation: the emission is then dominated by HMXBs 
(i.e., by accretion-powered X-ray pulsars) whose spectral slope $\gamma=
1.2$ (White et al. 1983) is reflected in the USC spectral parameter $\Gamma 
\simeq 1.2$.}
\end{figure}

For M82 and N253 the USC fluxes, $f_{\rm 2.1-10\, keV}=2.8 \times 10^{-11}$ 
and $4.6 \times 10^{-12}$ erg cm$^{-2}$ s$^{-1}$, respectively (see Fig.4) 
-- and corresponding luminosities $L_{\rm 2.1-10\, keV}=2.4 \times 10^{40}$ 
and $6 \times 10^{39}$ erg s$^{-1}$ -- imply populations of $\mincir 2.4 
\times 10^3$ and $\mincir 600$ luminous ($\magcir 10^{37}$ erg s$^{-1}$) 
X-ray binaries (if the hard components are totally produced by X-ray 
binaries). This means that these local SBGs host a factor $4-16$ times more 
high-luminosity X-ray binaries than the Galaxy: assuming a universal 
stellar IMF, a similar proportion holds between the respective star formation 
rates (a compatible result is based on radio SN rates, see Bartel et al. 
1987). This estimate is compatible with the hypothesis of an 
X-ray-binary--related origin of the hard component of SBG spectra.

\begin{figure}

\vspace{3.4cm}
\includegraphics{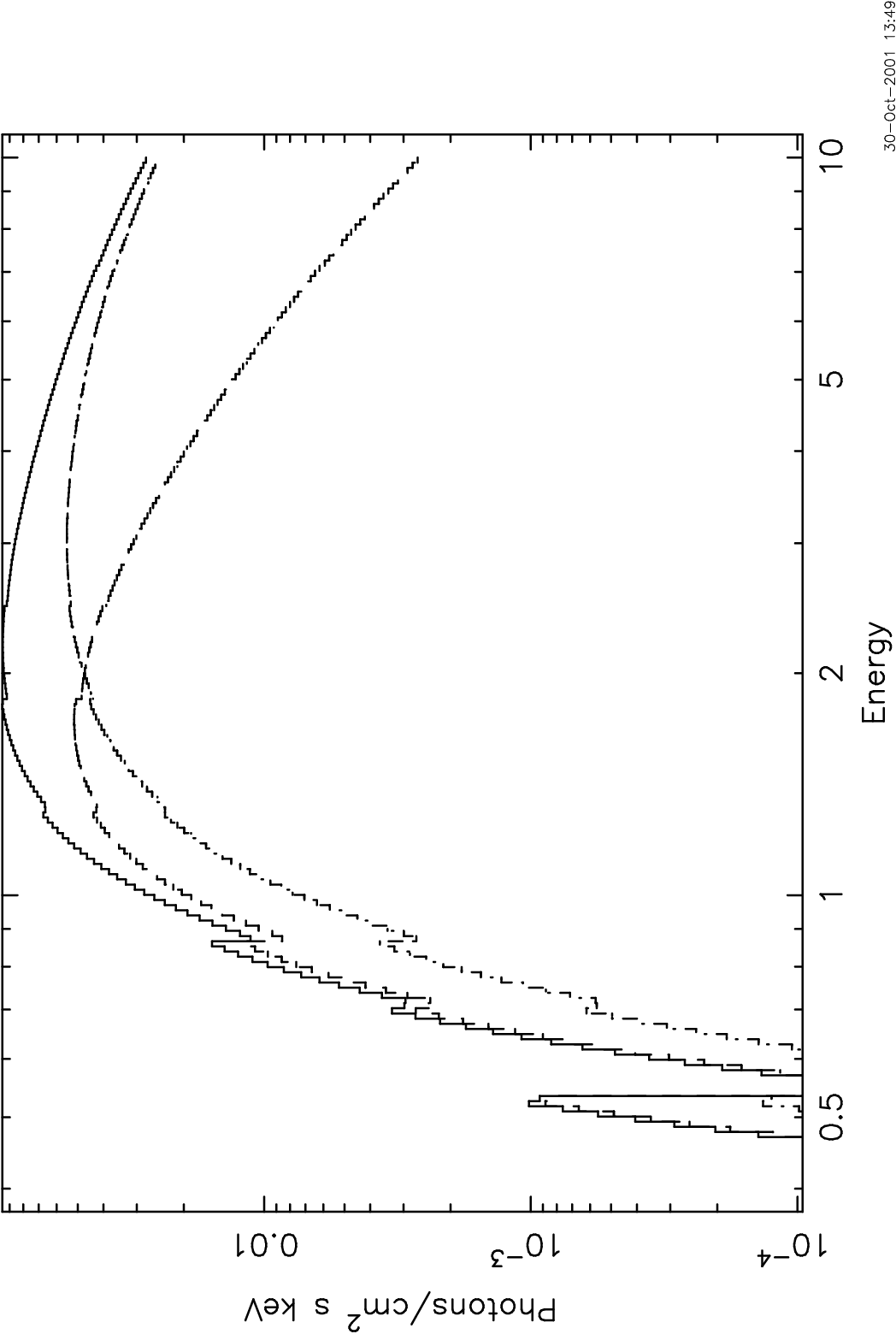}
\hspace{.5cm} 
\includegraphics{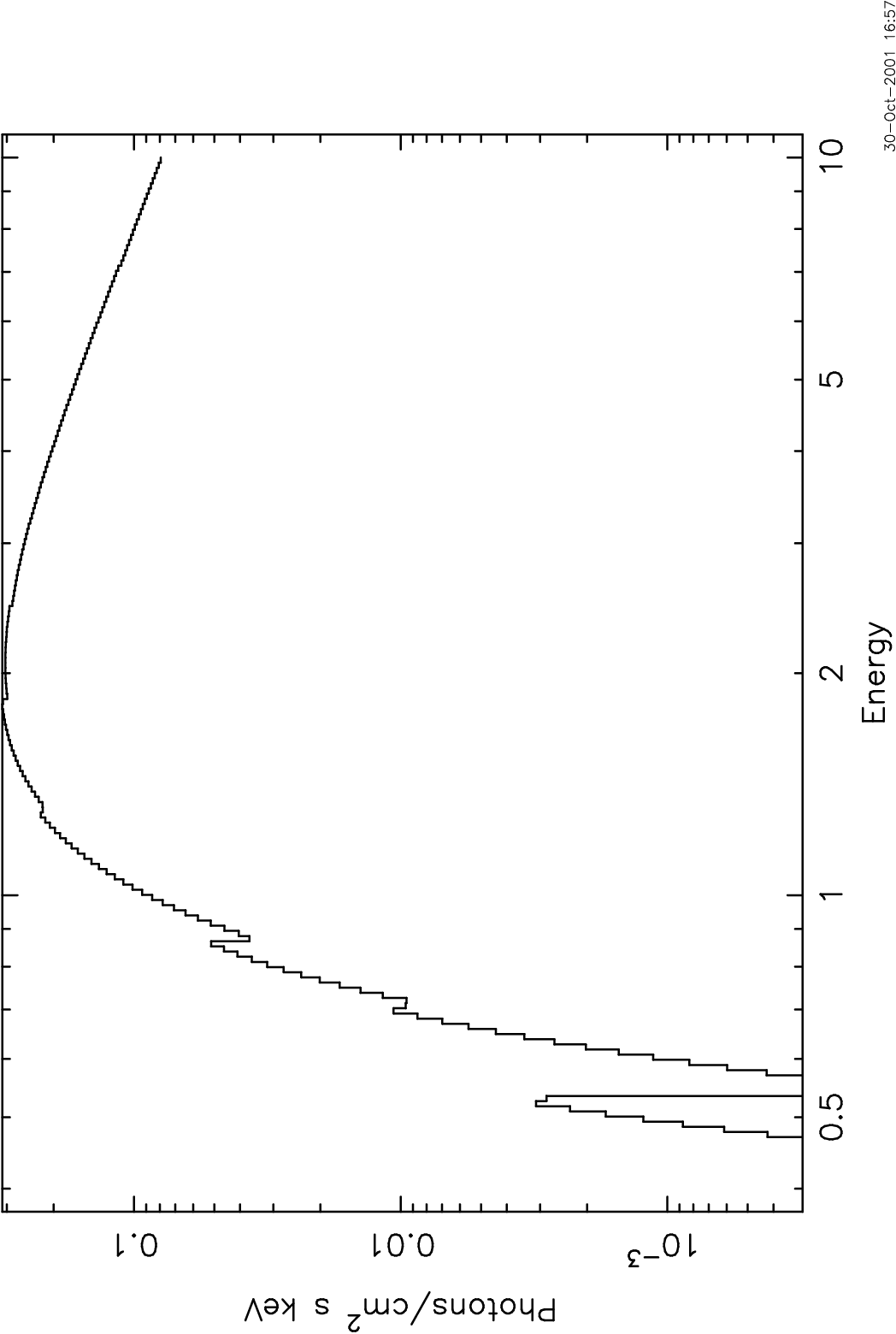}
\caption{The population-averaged LMXB spectrum (in ascending order: faint 
and bright systems, respectively, with relevant parameters and normalizations 
as in Fig.1; {\it left}), and a power-law spectrum $\propto \e^{-\gamma}$ 
with index $\gamma=1.1$ ({\it right}). Both spectra are absorbed through a 
hydrogen column density of $n_{\rm H}= 10^{22}$ cm$^{-2}$.}
\end{figure}

\section{Conclusion}

Using an equilibrium stellar-population synthesis model that follows 
the evolution of massive binary stars taking into account the different modes 
of mass transfer (Iben et al. 1995a,b), birthrates have been deduced for 
massive binaries, in which mass is accreted from a primary star onto a 
degenerate companion, for a range of primary masses. This, together with 
estimates of the duration of the X-ray bright phase, has led to realistic 
estimates of the relative abundances of HMXBs and LMXBs. A similar approach 
has been used for SNRs, following SN explosions that have occurred in both 
single and binary stars. 
The lifespans assumed for the three types of source ($\sim 2.5 \times 
10^4$ yr for HMXBs; $\mincir 10^7$ yr for LMXBs; and $\sim 10^3$ yr for 
SNRs -- see sections 2.1.1, 2.1.2, and 2.3) are safely shorter than a typical 
galactic SB lifetime ($\sim 10^8$ yr): this suggests that the equilibrium 
assumption implicit in our calculation is valid 
      \footnote{In our calculation we assume that, for each class of 
	sources, $\nu  \tau_{\rm x}$ systems have a spectrum with shape 
	$f(\e)$ and strength (i.e., integrated X-ray luminosity) $L_{\rm x}$. 
	This assumption is valid when the stellar population is in 
      equilibrium, i.e. when the age of the SB is much longer than the 
      characteristic lifespans of the systems.}.
For each source class, a systematic search of published spectra has shown 
that the average spectrum can be described as: (1) power law (with cutoff) 
for HMXBs, (2) unsaturated Comptonization model (with a correlation between 
shape and luminosity) for LMXBs, and (3) thermal plasma for SNRs. 

>From the spectral properties and relative abundances of the various classes 
of stellar sources, we have then computed the composite X-ray spectrum 
arising from a stellar population of Galactic composition. This 
"stellar" contribution has no essential degrees of freedom: it is fixed by 
the synthetic model and the observed X-ray spectra of the contributing 
components. 

The extended, non-stellar part of the emission has both thermal and 
non-thermal components. The former originates mostly from regions of 
interaction between the outgoing galactic wind powered by SN explosions and 
the ambient ISM. Non-thermal emission is due to Compton scattering of the 
SN-accelerated, radio-emitting relativistic electrons off the FIR and CMB 
radiation fields; the integrated spectrum is power-law, with index that 
should be comparable to that of the extended radio emission. The main 
quantities in the modelling of the diffuse emission are: (1) the gas/stars 
mass ratio, (2) the chemical abundance of the gas, and (3) the FIR radiation 
field (for a {\it given} electron energy spectrum, as deduced from the 
observed radio emission). 

This systematic survey of sources of X-ray emission in a SBG reveals that 
-- based on stellar evolution arguments -- in the 2-10 keV energy range 
X-ray binaries (1) are the most prominent stellar component, and (2) have 
the the required spectral shape. Their population-averaged spectrum is 
effectively described by a cut-off power law (plus an additional Fe-K 
emission, from HMXBs), whose observed ranges of parameter values 
provide good fits to the {\it BeppoSAX} spectra of the most extensively 
observed nearby SBGs, M82 and N253. This agreement also suggests that the 
hard component observed in the 2-10 keV spectrum of SBGs may indeed 
result from the integrated emission of X-ray binaries with a mean spectrum 
similar to the one observed from Galactic HMXBs or lower-luminosity 
LMXBs; or that some level of steeper Compton emission combines with the 
predicted "stellar" emission to give the observed hard component. In 
possible agreement with this suggestion is the conclusion, based on {\it 
Chandra} data, that in both N253 and M82 the $\sim$ 2-10 keV flux is 
dominated by (previously unresolved) point-source emission (Strickland et 
al. 2000, Griffiths et al. 2000). If our composite SBG spectrum, with its 
main components as identified in this paper, will be found successful also 
in fitting other SBG spectra (Persic et al., in preparation), then the 
central role played by X-ray binaries in SBGs will have been broadly 
demonstrated and ascertained.

In a broader perspective, our proposed template spectrum provides a physical 
fit that may prove especially useful for interpreting low spatial resolution 
data on normal and starbursting galaxies, either distant (e.g., from {\it 
Chandra} and {\it XMM} deep surveys) or nearby (e.g., from {\it ASCA} and 
{\it BeppoSAX}). Concerning the latter, the ambiguity noted so far in 
interpreting the available low-resolution spectra of SBGs -- notably, on the 
nature of the hard component (e.g.: Ptak et al. 1997, Okada et al. 1997, 
Persic et al. 1998, Cappi et al. 1999, Della Ceca et al. 1999, Moran et al. 
1999, Zezas et al. 1998; see also Weaver et al. 2000 and Dahlem et al. 2000) 
-- may disappear when a more realistic model is used to interpret the 
emission. One further consequence of using this template spectrum is the 
general possibility, for the first time, of measuring the star formation 
rate in galaxies directly from X-ray spectra. In fact, provided that the 
stellar IMF as well as the formation mechanisms (and hence the X-ray spectral 
properties) of X-ray binaries and SN are universal, in matching the 
observed spectrum of a given galaxy with a spectral template the only basic 
degree of freedom left will be the amplitude (i.e., the flux and hence the 
luminosity, once the distance to the source is known) which is proportional 
to the galaxy's star formation rate.

\vglue 0.3truecm
\noindent
{\it Acknowledgements.} 
We thank Massimo Cappi, John Danziger, Duane Gruber, Una Hwang, Elena Pian, 
and Jean Swank for useful exchanges and help. Two independent anonymous 
referees made very good suggestions. MP gratefully acknowledges financial 
support from the Italian {\sl Ministero dell'Universit\`a e della Ricerca 
Scientifica e Tecnologica} through grant No.9802192401$_{-}$007, and 
acknowledges the hospitality of the Center for Astrophysics and Space 
Sciences of the University of California, San Diego, where part of this 
work was carried out.
\vglue 0.5truecm

\def\ref{\par\noindent\hangindent 20pt} 

\noindent 
{\bf References} 
\vglue 0.2truecm 

\ref{Balzano, V.A. 1983, ApJ, 268, 602}
\ref{Bartel, N., Ratner, M.I., Shapiro, I.I., Rogers, A.E.E., \& Preston, R.A. 1987, ApJ, 323, 505}
\ref{Bocchino, F., Vink, J., Favata, F., Maggio, A., \& Sciortino, S. 2000, A\&A, 360, 671}
\ref{Boldt, E.A., Holt, S.S., Rothschild, R.E., \& Serlemitsos, P.J. 1976, A\&A, 50, 161}
\ref{Bondi, H., \& Hoyle, F. 1944, MNRAS, 104, 273}
\ref{Borkowski, K.J., Lyerly, W.J., \& Reynolds, S.P. 2001a, ApJ, 548, 820}
\ref{Borkowski, K.J., Rho, J., Reynolds, S.P., \& Dyer, K.K. 2001b, ApJ, 550, 334}
\ref{Burrows, D.N., Michael, E., Hwang, U., McCray, R., Chevalier, R.A., Petre, R., Garmire, G.P., Holt, S.S., \& 
     Nousek, J.A. 2000, ApJ, 543, L149}
\ref{Callanan, P.J., Penny, A.J., \& Charles, P.A. 1995, MNRAS 273, 201}
\ref{Cappi, M., Persic, M., Bassani, L., Franceschini, A., Hunt, L.K., Molendi, S., Palazzi, E., Palumbo, G.G.C.,
     Rephaeli, Y., \& Salucci, P. 1999, A\&A, 350, 777}
\ref{Cervi\~no, M., \& Mas-Hesse, J.M. 1994, A\&A, 284, 749}
\ref{Charles, P.A., \& Seward, F.D. 1995, in "Exploring the X-ray Universe" (Cambridge University Press)}
\ref{Chevalier, R.A. 1977, ARAA, 15, 175}
\ref{Chiosi, C., \& Maeder, A. 1986, ARAA, 24, 329}
\ref{Christian, D.J., \& Swank, J.H. 1997, ApJS, 109, 177}
\ref{Cid-Fernandes, R., Plewa, T., Rózyczka, M., Franco, J., Terlevich, R., Tenorio-Tagle, G., \& W.Miller 1996,
     MNRAS, 283, 419}
\ref{Collura, A., Reale, F., Schulman, E., \& Bregman, J.N. 1994, ApJ, 420, L63}
\ref{Condon, J.J., Helou, G., Sanders, D.B., \& Soifer, B.T. 1990, ApJS, 73, 359}
\ref{Condon, J.J., Huang, Z.-P., Yin, Q.F., \& Thuan, T.X. 1991, ApJ, 378, 65}
\ref{Conti, P.S., \& Underhill, A.B. 1988, O-Type Stars and Wolf-Rayet Stars, NASA, SP-497}
\ref{Corcoran, M.F., Waldron, W.L., MacFarlane, J.J., Chen, W., Pollock, A.M.T., Torii, K., Kitamoto, S., 
     Miura, N., Egoshi, M., \& Ohno, Y. 1994, ApJ, L95}
\ref{Dahlem, M., Heckman, T.M., \& Fabbiano, G. 1995, ApJ, 442, L49}
\ref{Dahlem, M., Parmar, A., Oosterbroek, T., Orr, A., Weaver, K.A., \& Heckman, T.M. 2000, ApJ, 538, 555}
\ref{Dalton, W.W., \& Sarazin, C.L. 1995a, ApJ, 440, 280}
\ref{Dalton, W.W., \& Sarazin, C.L. 1995b, ApJ, 448, 369}
\ref{David, L.P., Jones, C., \& Forman, W. 1992, ApJ, 388, 82}
\ref{Della Ceca, R., Griffiths, R.E., Heckman, T.M., Lehnert, M.D., \& Weaver, K.A. 1999, ApJ, 514, 772}
\ref{Della Ceca, R., Pellegrini, S., Bassani, L., Beckmann, V., Cappi, M., Palumbo, G., Trinchieri, G., \& 
     Wolter, A. 2001, A\&A, 375, 781}
\ref{Dewey, R.J., \& Cordes, J.M. 1987, ApJ, 321, 780}
\ref{Doane, J.S., \& Mathews, W.G. 1993, ApJ, 419, 573}
\ref{Ebisawa, K., Day, C.S.R., Kallman, T.R., Nagase, F., Kotani, T., Kawashima, K., Kitamoto, S., \& 
     Woo, J.W. 1996, PASJ, 48, 425}
\ref{Ebisawa, K., Ogawa, M., Aoki, T., Dotani, T., Takizawa, M., Tanaka, Y., \& Yoshida, K. 1994, PASJ, 46, 375}
\ref{Fabbiano, G. 1995, in "X-Ray Binaries", ed. W.H.G. Lewin, J. van Paradijs, \& 
     E.P.S. van den Heuvel (Cambridge University Press), 390}
\ref{Frail, D.A., Kulkarni, S.R., Sari, R., Djorgovski, S.G., Bloom, J.S., Galama, T.J., 
     Reichart, D.E., Berger, E., Harrison, F.A., Price, P.A., Yost, S.A., Diercks, A., 
     Goodrich, R.W., \& Chaffee, F. 2001, Nature, submitted (astro-ph/0102282)}
\ref{Franceschini, A., Bassani, L., Cappi, M., Granato, G.L., Malaguti, G., Palazzi, E., \& Persic, M. 2000, 
     A\&A, 353, 910}
\ref{Garc{\'\i}a-Vargas, M.L., Bressan, A., \& D{\'\i}az, A.L. 1995a, A\&AS, 112, 13}
\ref{Garc{\'\i}a-Vargas, M.L., Bressan, A., \& D{\'\i}az, A.L. 1995b, A\&AS, 112, 35}
\ref{Garmany, C.D., Conti, P.S., \& Massey, P. 1980, ApJ, 242, 1063}
\ref{Garwood, R.W., Dickey, J.M., \& Helou, G. 1987, ApJ, 322, 88}
\ref{Goldshmidt, O., \& Rephaeli, Y. 1995, ApJ, 444, 113}
\ref{Griffiths, R.E., Ptak, A., Feigelson, E.D. Garmire, G., Townsley, L., Brandt, W.N., Sambruna, R., \& 
     Bregman, J.N. 2000, Science, 290, 1325}
\ref{Hamilton, A.J.S., \& Sarazin, C.L. 1984, ApJ, 284, 601}
\ref{Hickson, P., Menon, T.K., Palumbo, G.G.C., \& Persic, M. 1989, ApJ, 341, 679}
\ref{Hwang, U. 2001, in "Young Supernova Remnants", eds. S.S. Holt \& U. Hwang, AIP Conference Proceedings 
     Vol. 565, 143}
\ref{Iben, I., Jr., Tutukov, A.V., \& Yungelson, L.R. 1995a, ApJS, 100, 217}
\ref{Iben, I., Jr., Tutukov, A.V., \& Yungelson, L.R. 1995b, ApJS, 100, 233}
\ref{Ishisaki, Y., Makishima, K., Iyomoto, N., Hayashida, K., Inoue, H., Mitsuda, K., Tanaka, Y., Uno, S., 
     Kohmura, Y., Mushotzky, R.F., Petre, R., Serlemitsos, P.J., \& Terashima, Y. 1996, PASJ, 48, 237}
\ref{J{\o}rgensen, H., Lipunov, V.M., Panchenko, I.E., Postnov, K.A., \& Prokhorov, M.E. 1997, ApJ, 486, 110}
\ref{Kaaret, P., Prestwich, A.H., Zezas, A., Murray, S.S., Kim, D.-W., Kilgard, R.E., Schlegel, E.M., \& 
     Ward, M.J. 2001, MNRAS, 321, L29}
\ref{Keil, R., Boller, Th., \& Fujimoto, R. 2001, in "New Century of X-Ray Astronomy", eds. H. Kunieda \& H. Inoue, 
     ASP Conference Series, in press (astro-ph/0106195)}
\ref{King, A.R., Davies, M.B., Ward, M.J., Fabbiano, G., \& Elvis, M. 2001, ApJ, 552, L109}
\ref{Klein, U., Wielebinski, R., \& Morsi, H.W. 1988, A\&A, 190, 41}
\ref{Kouveliotou, C., van Paradijs, J., Fishman, G.J., Briggs, M.S., Kommers, J., Harmon, B.A., Meegan, C.A., 
     \& Lewin, W.H.G. 1996, Nature, 379, 799}
\ref{Kronberg, P.P., Biermann, P.L., \& Schwab, F.R. 1985, ApJ, 291, 693}
\ref{Leitherer, C., \& Heckman, T.M. 1995, ApJS, 96, 9}
\ref{Levenson, N.A., Weaver, K.A., \& Heckman, T.M. 2001a, ApJS, 133, 269}
\ref{Levenson, N.A., Weaver, K.A., \& Heckman, T.M. 2001b, ApJ, 550, 230}
\ref{Lightman, A.P., \& White, T.R. 1988, ApJ, 335, 57}
\ref{Lipunov, V.M., Ozernoy, L.M., Popov, S.B., Postnov, K.A., \& Prokhorov, M.E. 1996a, ApJ, 466, 234}
\ref{Lipunov, V.M., Postnov, K.A., \& Prokhorov, M.E. 1996b, A\&A, 310, 489}
\ref{Mac Fadyen, A.I., \& Woosley, S.E. 1999, ApJ, 524, 262}
\ref{Madau, P., Ferguson, H.C., Dickinson, M.E., Giavalisco, M., Steidel, C.C., \& Fruchter, A. 1996, MNRAS, 283, 1388}
\ref{Maeda, Y., Baganoff, F.K., Feigelson, E.D., Morris, M., Bautz, M.W., Brandt, W.N., Burrows, D.N., Doty, J.P., 
     Garmire, G.P., Pravdo, S.H., Ricker, G.R., \& Townsley, L.K. 2001, ApJ, in press (astro-ph/0102183)}
\ref{Maeder, A., \& Meynet, G. 1989, A\&A, 210, 155}
\ref{Magdziarz, P., \& Zdziarski, A.A. 1995, MNRAS, 273, 837}
\ref{Mas-Hesse, J.M., \& Cervi\~no, M. 1999, in IAU Symp. 193 "WR Phenomena in Massive Stars and Starburst Galaxies", 
     eds. K.A. Van der Hucht, G. Koenigsberger, \& P.R.J. Eenens (San Francisco, CA: Astronomical Society of the 
     Pacific), 550}
\ref{Mas-Hesse, J.M., Cervi\~no, M., Rodr{\'\i}guez-Pascual, P.M., \& Boller, Th. 1996, A\&A, 309, 431}
\ref{Mas-Hesse, J.M., \& Kunth, D. 1991, A\&AS, 88, 399}
\ref{Matsumoto, H., Tsuru, T.G., Koyama, K., Awaki, H., Canizares, C.R., Kawai, N., Matsushita, S., \& 
     Kawabe, R. 2001, ApJ, 547, L25}
\ref{Matsushita, K., Makishima, K., Awaki, H., Canizares, C.R., Fabian, A.C., Fukazawa, Y., Loewenstein, M., 
     Matsumoto, H., Mihara, T., Mushotzky, R.F., Ohashi, T., Ricker, G.R., Serlemitsos, P.J., Tsuru, T., 
     Tsusaka, Y., \& Yamazaki, T. 1994, ApJ, 436, L41}
\ref{Matsushita, K., Ohashi, T., \& Makishima, K. 2000, PASJ, 52, 685}
\ref{Meynet, G. 1995, A\&A, 298, 767}
\ref{Meurs, E.J., \& van der Heuvel, E.P. 1989, A\&A, 226, 88}
\ref{Mihara, T., Makishima, K., Kamijo, S., Ohashi, T., Nagase, F., Tanaka, Y., \& Koyama, K. 1991, 
     ApJ, 379, L61}
\ref{Moran, E.C., Lehnert, M.D., \& Helfand, D.J. 1999, ApJ, 526, 649}
\ref{Nagase, F., Zylstra, G., Sonobe, T., Kotani, T., Inoue, H., \& Woo, J. 1994, ApJ, 436, L1}
\ref{O'Connell, R.W., Gallagher, J.S. III, Hunter, D.A., \& Colley, W.N. 1995, ApJ, 446, L1}
\ref{Okada, K., Mitsuda, K., \& Dotani, T. 1997, PASJ, 49, 653}
\ref{Ostriker, J.E. 1977, Ann. N.Y. Ac. Sci., 302, 229}
\ref{Paciesas, W.S., Meegan, C.A., Pendleton, G.N., Briggs, M.S., Kouveliotou, C., Koshut, T.M., 
     Lestrade, J.P., McCollough, M.L., Brainerd, J.J., Hakkila, J., Henze, W., Preece, R.D., 
     Connaughton, V., Kippen, R.M., Mallozzi, R.S., Fishman, G.J., Richardson, G.A., \& Sahi, M. 
     1999, ApJS, 122, 465}
\ref{Pence, W.D., Snowden, S.L., Mukai, K., \& Kuntz, K.D. 2001, ApJ, 561, 189}
\ref{P\'erez-Olea, D.E., \& Colina, L. 1995, MNRAS, 277, 857}
\ref{Persic, M., Mariani, S., Cappi, Bassani, L., Danese, L., Dean, A.J., Di Cocco, D., Franceschini, A., 
     Hunt, L.K., Matteucci, F., Palazzi, E., Palumbo, G.G.C., Rephaeli, Y., Salucci, P., \& Spizzichino, A. 
     1998, A\&A, 339, L33}
\ref{Pietsch, W., Roberts, T.P., Sako, M., Freyberg, M.J., Read, A.M., Borozdin, K.N., Branduardi-Raymont, G., 
     Cappi, M., Ehle, M., Ferrando, P., Kahn, S.M., Ponman, T.J., Ptak, A., Shirey, R.E., \& Ward, M. 2001, 
     A\&A, 365, L174}
\ref{Piran, T. 1999, Phys Rep, 314, 575}
\ref{Pols, O.R., Coté, J., Waters, L.B.F.M., \& Heise, J. 1991, A\&A, 241, 119}
\ref{Pols, O.R., \& Marinus, M. 1994, A\&A, 288, 475}
\ref{Pravdo, S.H., Becker, R.H., Boldt, E.A., Holt, S.S., Serlemitsos, P.J., \& Swank, J.H. 1977, ApJ, 215, L61}
\ref{Ptak, A., Serlemitsos, P.J., Yaqoob, T., Mushotzky, R., \& Tsuru, T. 1997, AJ, 113, 1286}
\ref{Ptak, A., \& Griffiths, R. 1999, ApJ, 517, L85}
\ref{Recchi, S., Matteucci, F., \& D'Ercole, A. 2001, MNRAS, 322, 800}
\ref{Rephaeli, Y. 1979, ApJ, 227, 364}
\ref{Rephaeli, Y., \& Gruber, D. 2001, submitted to AA}
\ref{Rephaeli, Y., Gruber, D., Persic, M., \& MacDonald, D., 1991, ApJ, 380, L59}
\ref{Rephaeli, Y., Gruber, D., \& Persic, M., 1995, AA, 300, 91}
\ref{Rieke, G.H., Loken, K., Rieke, M.J., \& Tamblyn, P. 1993, ApJ, 412, 99}
\ref{Roberts, T.R., \& Warwick, R.S. 2000, MNRAS, 315, 98}
\ref{Rosner, R., Golub, L., \& Vaiana, G.S. 1985, ARA\&A, 23, 413}
\ref{Sako, M., Liedahl, D.A., Kahn, S.M., \& Paerels, F. 1999, ApJ, 525, 921}
\ref{Sari, R., Piran, T., \& Narayan, R. 1998, ApJL, 497, L17}
\ref{Schaerer, D., \& Vacca, W.D. 1998, ApJ, 497, 618}
\ref{Schmitt, H.R., Kinney, A.L., Calzetti, D., \& Storchi Bergmann, T. 1997, AJ, 114, 592}
\ref{Searle, L., Sargent, W.L.W., \& Bagnuolo, W.G. 1973, ApJ, 179, 427}
\ref{Soifer, B.T., Sanders, D.B., Neugebauer, G., Danielson, G.E., Lonsdale, C.J., Madore, B.F., \& 
     Persson, S.E. 1986, ApJ, 303, L41}
\ref{Slane, P., Hughes, J.P., Edgar, R.J., Plucinsky, P.P., Miyata, E., Tsunemi, H., \& Aschenbach, B. 2001, 
     ApJ, 548, 814}
\ref{Strickland, D.K., Colbert, E.J.M., Heckman, T.M., Weaver, K.A., Dahlem, M., \& Stevens, I.R. 
     2001, ApJ, 560, 707}
\ref{Strickland, D.K., Heckman, T.M., Weaver, K.A., \& Dahlem, M. 2000, AJ, 120, 2965}
\ref{Strickland, D.K., \& Stevens, I.R. 2000, MNRAS, 314, 511}
\ref{Suchkov, A.A., Balsara, D.S., Heckman, T.M., \& Leitherner, C. 1994, ApJ, 430, 511}
\ref{Sunyaev, R.A., \& Titarchuk, L.G. 1980, A\&A, 86, 121}
\ref{Tanaka, Y., \& Lewin, W.H.G. 1995, in "X-Ray Binaries", ed. W.H.G. Lewin, J. van Paradijs, 
     \& E.P.J. van den Heuvel (Cambridge: Cambridge University Press), 126}
\ref{Terlevich, R., Tenorio-Tagle, G., Franco, J., \& Melnick, J. 1992, MNRAS, 255, 713}
\ref{Thompson, R.I., Weymann, R.J., \& Storrie-Lombardi, L.J. 2001, ApJ, 546, 694}
\ref{Treves, A., Belloni, T., Chiappetti, L., Maraschi, L., Stella, L., Tanzi, E.G., \& van der Klis, M. 1988, 
     ApJ, 325, 119}
\ref{Trinchieri, G., Israel, G.L., Chiappetti, L., Belloni, T., Stella, L., Primini, F., Fabbiano, G., \& 
     Pietsch, W. 1999, A\&A, 348, 43}
\ref{Tsuru, T., Awaki, H., Koyama, K., \& Ptak, A. 1997, PASJ, 49, 619}
\ref{Tutukov, A.V., Yungelson, L.R., \& Iben, I. 1992, ApJ, 386, 197}
\ref{Van Bever, J., Belkus, H., Vanbeveren, D., \& Van Rosenbergen, W. 1999, New Astronomy, 4, 173}
\ref{Van Bever, J., \& Vanbeveren, D. 1998, A\&A, 334, 21}
\ref{Van Bever, J., \& Vanbeveren, D. 2000, A\&A, 358, 462}
\ref{Vanbeveren, D., Van Bever, J., \& De Donder, E. 1997, A\&A, 317, 487}
\ref{Vanbeveren, D., De Donder, E., Van Bever, J., \& Van Rensbergen, W., \& De Loore, C. 1998,
     New Astronomy, 3, 443}
\ref{van Paradijs, J. 1998, in "The Many Faces of Neutron Stars", ed. R.Buccheri et al. (Kluwer 
     Academic Publishers), 279}
\ref{van Paradijs, J., Kouveliotou, C., \& Wijers, R.A.M.J. 2000, ARAA, 38, 379}
\ref{Veilleux, S. 2000, in "Starbursts -- Near and Far", ed. Tacconi, L., \& Lutz, D. (Springer), 
     in press (astro-ph/0012121)}
\ref{Vink, J., Kaastra, J.S., Bleeker, J.A.M., Preite-Martinez, A. 2000, A\&A, 354, 931}
\ref{Wang, Z., \& Helou, G. 1992, ApJ, 398, L33}
\ref{Waters, L.B., \& van Kerkwijk, M.H. 1989, ApJ, 223, 196}
\ref{Weaver, K.A., Heckman, T., \& Dahlem, M. 2000, ApJ, 534, 684}
\ref{Weedman, D.W., Feldman, F.R., Balzano, V.A., Ramsey, L.W., Sramek, R.A., \& Wuu, C.-C. 1981, ApJ, 248, 105}
\ref{White, N.E., \& Marshall, F.E. 1984, ApJ, 281, 354}
\ref{White, N.E., Nagase, F., \& Parmar, A.N. 1995, in "X-Ray Binaries", ed. W.H.G. Lewin, J. van Paradijs, \& 
     E.P.S. van den Heuvel (Cambridge University Press), 1}
\ref{White, N.E., Peacock, A., \& Taylor, B.G. 1985, ApJ, 296, 475} 
\ref{White, N.E., Stella, L., \& Parmar, A.N. 1988, ApJ, 324, 363}
\ref{White, N.E., Swank, J.H., \& Holt, S.S. 1983, ApJ, 270, 711}
\ref{White, N.E., \& van Paradijs, J. 1996, ApJ, 473, L25}
\ref{Wilson, C.K., \& Rothschild, R. 1983, ApJ, 274, 717}
\ref{Woltjer, L. 1972, ARAA, 10, 129}
\ref{Woosley, S.E., \& Weaver, T.A. 1995, ApJS, 101, 181}
\ref{Wu, K. 2001, PASA, in press (astro-ph/0103157)}
\ref{Wu, K., Tennant, A., Swartz, D., \& Ghosh, K. 2001, ApJ, submitted}
\ref{Wunderlich, E., \& Klein, U. 1988, A\&A, 206, 47}
\ref{Zezas, A.L., Georgantopoulos, I., \& Ward, M.J. 1998, MNRAS, 301, 915}
\vglue 1.0truecm

\end{document}